%% file: mass.tex
\documentclass{cernyrep}
\usepackage{graphicx}
\usepackage{texnames}
\usepackage[T1]{fontenc}
\usepackage{amssymb}
\usepackage{cernunits}
\usepackage{psfrag}
\include{cernunits.sty}
\include{wasyfont}
\begin{document}
\title{Mass Issues in Fundamental Interactions\footnote{\quad Based on three lectures given at the 2008 European School of High-Energy Physics (Herbeumont, Belgium,  8 June -- 21 June 2008).}}
 
\author{Jean-Marc G\'erard}

\institute{Centre for Particle Physics and Phenomenology (CP3), \\ Universit\'e catholique de Louvain, Belgium}

\maketitle 

\begin{abstract}
Driven by the mass problem, we raise some issues of the fundamental interactions in terms of non trivial commutation relations implemented within toy theories.
\end{abstract}
 
\section*{Introduction}
 
 The four known basic forces of nature turn out to proceed from a universal gauge principle. In particular, Einstein's general theory of relativity can be consi\-dered as the first Yang-Mills theory. Indeed photons do not carry an electric charge but gravitons appear to gravitate the way gluons glue in Quantum Chromo-Dynamics. The confinement and spontaneous symmetry breaking mechanisms put forward to prevent long range nuclear forces form nowadays the cornerstone of the Standard Model for particle physics. Such subtle issues to get round gauge invariance are highly suspected to be responsible for an explicit violation of the invariance under time-reversal in strong and weak interactions, respectively.  
 
\par\noindent
The following three Sections are built upon the problem of mass.

\begin{itemize}
\item In the first Section, a geometrical interpretation of non-abelian gauge invariance is outlined from the striking fact that somebody in a free-falling elevator would experience no apparent weight. Our main goal here is to display how universal the basic forces may be, through the concepts of mass and energy. For that purpose, we mostly rely on a  \textit{scalar}  theory for gravity which allows us to elude tedious tensor calculus.  
\item  In the second Section, we make use of an effective theory for strong interactions to explain the origin of nucleon masses. We limit ourselves to the case of  \textit{two light flavours} and emphasize that the observed proton-neutron mass splitting might imply a large electric dipole moment for the neutron.
\item In the last Section, inspired by the chiral symmetry breaking at work in the theory for strong interactions, we consider an effective theory for electroweak interactions to explain the origin of boson and fermion masses. We illustrate how Yukawa interactions allow in principle a matter-antimatter asymmetry,  \textit{regardless of the flavour mixing pattern}.
\end{itemize}

\section{Gauge Invariance and \mbox{\boldmath $ [D_\mu \ , D_\nu] \neq 0$}}

\subsection{Weight of compact bodies}

The beauty of modern physics lies in the fact that it allows us not only to relate seemingly different phenomena, such as the fall of a ripe apple and the motion of the full moon or electricity and magnetism, but also to unify apparently independent everyday concepts such as rest and uniform motion, space and time or gravitation and acceleration. In this way, we now have at our disposal a well-defined theoretical frame to explain why we do not feel the gravitational field of the Sun, but also to formulate rather precise questions about the origin of our weight, at least within the precision of a usual bathroom scales...

\par\noindent
Our weight is obviously contingent upon the gravitational force exerted by the Earth:
\[
\vec{W} = m_{gr} \   \vec{g} \qquad  (g = \frac{GM_\oplus}{R_\oplus^2})
\eqno{(1.1)}
\]
and its precise value depends on our location (altitude but also latitude). As opposed to weight, mass appears to be an intrinsic property of matter which relates its manifest response (acceleration) to an abstract cause (force) in classical mechanics:
\[
\vec{F} = m_{in} \ \vec{a}.
\eqno{(1.2)}
\]
Stevin's drops from the top of a tower, Galileo's observations of wooden balls rolling down sloping planes and Newton's experiments with pendulums made of various materials indicated that all bodies tend to fall with the same acceleration at the surface of the Earth, no matter what their constitution may be, i.e.,
\[
m_{gr} = m_{in},
\eqno{(1.3)}
\]
with an accuracy of about  $10^{-3}$. More accurate torsion balance experiments initiated by the Hungarian Baron Roland von E\"otv\"os around \textbf{1890} nicely confirmed such a correlation between gravity and inertia at the level of $10^{-9}$. Nowadays, this equality between gravitational and inertial masses is firmly established at the level of  $10^{-12}$. The so-called "weak" equivalence principle rests upon Eq. (1.3).

\bigskip

\par\noindent
From the striking universality of free-fall (see the apple and the Moon falling towards the Earth)  Einstein inferred, as far back as \textbf{1907}, that his law which links mass to rest energy, i.e.,
\[
m = \frac{E_0}{c^2},
\eqno{(1.4)}
\]
 "holds not only for inertial but also for gravitational mass" \cite{1}. In other words, \textit{energy weighs}!  So, electromagnetic binding energies do equally contribute to the inertial and gravitational mass  such that all atoms (H, H$^\ast$,   $\overline{\textrm{H}}$,...) fall with the same acceleration. In particular, matter and antimatter  fall the same way since both represent positive energies.  The amazing   accuracy of modern experiments extends this "Einstein" equivalence principle to strong and weak nuclear binding energies since atoms are made of protons, neutrons and electrons. But what about gravitational bound states?

\par\noindent
For an homogeneous and spherical distribution of matter, the gravitational binding energy  
\[
\Omega \equiv -  \frac{1}{2}  \sum_{i,j} \  G \frac{m_i \  m_j}{r_{ij}}
\eqno{(1.5)}
\]
is simply given by
\[
\Omega = -  \frac{3}{5}  \frac{GM^2}{R}.
\eqno{(1.6)}
\]
From the magic relation between the Newton constant $G$,  the light velocity $c$ and the solar mass   $M_\odot \approx 2\times\!10^{30}$ kg,
\[
\frac{2 G  M_\odot}{c^2} \approx 3~\textrm{km}
\eqno{(1.7)}
\]
which warns you that the Sun confined inside a (Schwarzschild) radius of 3 km would simply be a black hole, we get a ratio of the internal gravitational binding energy to the total mass energy scaling like 
\[
s \equiv |\frac{\Omega}{Mc^2}| \approx \frac{3}{10} (\frac{3~\textrm{km}}{R}) (\frac{M}{2\times\!10^{30}~\textrm{kg}}).
\eqno{(1.8)}
\]
For a typical ball (say, $R$ = 10 cm, $M$ = 2 kg) we obtain in this manner a "sensitivity" (or compactness factor) of the order of  $10^{-26}$. Consequently, present E\"otv\"os-like laboratory experiments are totally unable to tell us whether the gravitational binding energy equally contributes to the inertial and to the gravitational mass. Let us therefore define the mass ratio for gravitational bound states as follows:
\[
\frac{m_{gr}}{m_{in}} \equiv 1 + \eta  \frac{\Omega}{Mc^2}
\eqno{(1.9)}
\]
with $\eta$, a dimensionless parameter measuring any departure from universality for compact bodies in free-fall. For an homogeneous Earth ($R$ $\approx$ 6400~km, $M \approx$ $6\!\times\!10^{24}$~kg) and Moon ($R$ $\approx$ 1700~km, $M$ $\approx 7\times\!10^{22}$~kg), the compactness factors are roughly  $4\times\!10^{-10}$ and  $2\times\!10^{-11}$, respectively. The observational fact that the Moon's orbit around the Earth does not appear to be continuously polarized towards the Sun \cite{2}  guarantees that they both fall towards the Sun at equal rates with an accuracy of about $2\!\times\!10^{-13}$. From the relation
\[
|\frac{a_\oplus-a_{\leftmoon}}{g}| = \eta | \frac{\Omega_\oplus}{M_\oplus c^2} - \frac{\Omega_{\leftmoon}}{M_{\leftmoon} c^2} |,
\eqno{(1.10)}
\]
we infer that their gravitational binding energy equally contributes to the inertial and to the gravitational mass with an accuracy of about $5\times\!10^{-4}$. A more careful analysis, taking into account the inhomogeneous distribution of matter in the Earth and Moon, gives the range \cite{3}
\[
 |\eta^{\textrm{{\tiny exp}}}| = (4.0 \pm 4.3) \times 10^{-4}.
 \eqno{(1.11)}
\]
Let us raise this empirical fact at the level of a "strong" equivalence principle (SEP) which simply states that the free-fall of a compact body is also independent of its gravitational binding energy, i.e., 
\[
\eta_{\textrm{{\tiny SEP}}} \equiv 0.
\eqno{(1.12)}
\]
The SEP can be considered as a physical principle which limits the choice of our theory for gravitation among all possible metric theories one can construct.

\subsection{Mass versus energy in gravitational interactions}

The relativistic (Lorentz invariant) action for a free elementary particle reads
\[
S^{\textrm{{\tiny free}}} = \int\!\!\{-m_{in} \  c^2\} d\tau = \int\!\!\{-m_{in} c^2 + m_{in} \frac{v^2}{2} + \mathcal{O}(\frac{1}{c^2})\} dt.
\eqno{(1.13)}
\]
If this massive particle carries an electric charge $q$ and freely propagates in an electromagnetic vector field $A^\mu (t,\vec{x})$, then the action becomes
\[
S^{\textrm{{\tiny e.m.}}} = \int\!\!\{-m_{in} c^2 - \frac{q}{c} \  \frac{dx_\mu}{d\tau} \  A^\mu\}d\tau. 
\eqno{(1.14)}
\]
By straight analogy with the Coulomb potential $A^0$ in the static limit $(dx_0 = c$ $dt, d\vec{x} = 0)$, the trajectory of an elementary particle propagating in a scalar gravitational field $V(t,\vec{x})$ might simply be defined by
\[
S^{\textrm{{\tiny gr.}}} = \int\!\!\{-m_{in} \  c^2 - m_{gr}V\} d\tau.
\eqno{(1.15)}
\]
If the weak equivalence principle (1.3) applies, this action  can equivalently be written as
\[
S^{\textrm{{\tiny gr.}}} = \int\!\! - m_{in} \  c \  ds,
\eqno{(1.16)}
\]
$ds$ being the invariant distance (or arclength) given by
\[
ds^2 = (1+\frac{V}{c^2})^2 \eta_{\mu\nu} dx^\mu dx^\nu.
\eqno{(1.17)}
\]
Proposed by the Finnish physicist G. Nordstr\"om in \textbf{1913}, i.e., two years before the birth of general relativity \cite{4}, this background-dependent scalar theory is thus characterized by a specific, conformally flat, space-time defined by
\[
g_{\mu\nu} = (1 + \frac{V}{c^2})^2 \eta_{\mu\nu}.
\eqno{(1.18)}
\]
In other words the physical metric $g_{\mu\nu}(t,\vec{x})$ has only one degree of freedom, a scalar graviton field, the rest being fixed a priori by the flat Minkowski metric $\eta_{\mu\nu}$ which acts here as an absolute background in a way consistent with the Einstein equivalence principle. As a direct consequence, any massless particle plunged in this scalar gravitational field keeps on propagating along the light-cone  
\[
ds^2 \  \propto \ \eta_{\mu\nu} dx^\mu dx^\nu = 0.
\eqno{(1.19)}
\]
In particular, the massless scalar graviton itself does not feel gravity and the strong   equivalence principle (1.12) obviously holds true since the  gravitational binding energy does not interfere in the free-fall of a body.

\bigskip

\par\noindent
The Nordstr\"om's theory with its prior space-time geometry \cite{5} has been the first, mathematically consistent, theory resolving the clash between Newton's instantaneous gravity and Einstein's special relativity. However this theory  has in fact been definitely falsified no more than six years after its elaboration. Following Nordstr\"om, the massless photon does not gravitate either and there is thus no possible light-bending at the limb of the Sun, in "flat"  contradiction with the direct observations \cite{6} made by Dyson and Eddington during a total solar eclipse in \textbf{1919}. Yet, since it embodies the strong equivalence principle, we shall rely on this rather simple toy theory in which only \underline{mass} can feel the gravitational degree of freedom. For a more realistic theory where gravity couples to all kinds of \underline{energy} in a way also compa\-tible with the SEP, one should introduce a formalism which is free of any prior space-time geometry, i.e.,   background-independent.

\bigskip

\par\noindent
Inspired by Nordstr\"om's theory where the equivalence principle has simply been geometrized, let us assume the gravitational interactions of matter (and light) to be characterized by the universal coupling to a metric field. For a free massive particle, this simply amounts to substituting $g_{\mu\nu} (q)$ for the rigid Minkowski metric $\eta_{\mu\nu}$ in Eq. (1.13):
\[
c^2 d\tau^2 \to ds^2 = g_{\mu\nu} (q) dq^\mu dq^\nu.
\eqno{(1.20)}
\]
The relativistic principle of "maximal aging", originally set forth for twins, extends to curved space-time if a local inertial frame can be defined on every segment of the free-body world line.  In this case, the variational principle
\[
\delta \int\!\! ds = 0
\eqno{(1.21)}
\]
implies that the track $q^\mu (\lambda)$ of a free particle plunged in a given gravitational field is always the shortest path (or geodesic) of the curved space-time, regardless of its (inertial) mass. Setting $d\lambda = ds$ on the unvaried path \underline{after} all partial derivatives have been evaluated in the generalized Euler-Lagrange equations of motion, 
\[
\{\frac{d}{d\lambda} \frac{\partial}{\partial q'^{\rho}} - \frac{\partial}{\partial q^ \rho}\}\ \{g_{\mu\nu}(q)q'^\mu q'^\nu\}^{\frac{1}{2}} = 0,
\eqno{(1.22)}
\]
one easily obtains
\[
\frac{d^2 q^\sigma}{ds^2} + \Gamma^\sigma_{\ \mu\nu} \frac{dq^\mu}{ds} \frac{dq^\nu}{ds} =  0,
\eqno{(1.23)}
\]
where
\[
\Gamma^\sigma_{\ \mu\nu} \equiv  \frac{1}{2}  g^{\sigma\rho} (\partial_\nu g_{\mu\rho} + \partial_\mu g_{\rho\nu} - \partial_\rho g_{\mu\nu})
\eqno{(1.24)}
\]
are the Christoffel symbols, also known as the components of the (affine) connection.

\bigskip

\par\noindent
For illustration, let us consider the stationary, inhomogeneous gravitational field  
\[
V(r) = - \frac{GM}{r}
\eqno{(1.25)}
\]
induced by the Sun on the Earth which is 150 millions kilometres away. It is enough that the mixed space-time components of the $\Gamma$ connection obey the approximate relation
\[
\Gamma^i_{00} = \frac{1}{c^2} \delta^{ik} \partial_k V  +  \mathcal{O} (\frac{1}{c^4})
\eqno{(1.26)}
\]
in the weak field approximation
\[
|\frac{V}{c^2}| \approx  \frac{1.5~\textrm{km}}{150\times\!10^6~\textrm{km}} = 10^{-8} \ll 1
\eqno{(1.27)}
\]
to recover  the Newtonian equation of motion 
\[
\frac{d^2 \vec{q}}{dt^2} + \vec\nabla V \approx \vec 0.
\eqno{(1.28)}
\]
As a consequence, the weak equivalence principle is automatically implemented through the \textit{kinematics} of test particles (space-time tells small mass how to move), without any reference to the specific \textit{dynamics} of gravity (large mass tells space-time how to curve). In the particular case of the Nordstr\"om scalar theory, one has indeed the exact relation
\[
\Gamma^i_{\ 00} = \delta^{ik} \  \partial_k \ln (1+\frac{V}{c^2}).
\eqno{(1.29)}
\]
 So,  what then does privilege Einstein's non-linear field equations which are supposed to determine the geometry around the Sun as well as the dynamics of the whole Universe? Here, we would like to emphasize that the free-fall for compact bodies (i.e., bodies containing non-negligible gravitational binding energy, in contrast to test bodies) may give us a clue.

\bigskip

\par\noindent
"If a person falls freely he will not feel his own weight" \cite{7}. From this early "happiest though", Einstein inferred that all physical laws of special relativity (electromagnetism included) should remain valid in a sufficiently small free-falling laboratory to eventually establish his quite successful general theory of relativity, more than eight years later. The geodesic equations of motion we have derived in Eq. (1.23) nicely illustrate this remarkable property. Indeed they can be interpreted as a generalized Newton first law of classical mechanics in the presence of gravitational forces: 
 \[
 Dp^\sigma \equiv (\partial_\nu p^\sigma + \Gamma^\sigma_{\ \mu\nu} p^\mu) dq^\nu = 0  
 \eqno{(1.30)}
\]
with $p^\sigma \equiv m dq^\sigma/d\tau$, the relativistic 4-momentum of a test particle. In an inertial (free-falling) frame, the Christoffel symbols $\Gamma^\sigma_{\ \mu\nu}$ which are \underline{not} the components of a general coordinate tensor identically vanish and the reduced equations of motion 
 \[
(\frac{d^2 x^\sigma}{d\tau^2})\left|\right._{\Gamma\to 0} = 0
\eqno{(1.31)}
\]
remain covariant with respect to (linear) Lorentz transformations, in full agreement with Einstein's equi\-valence principle. Similarly, in the limit of non-relativistic velocities the proper-time interval $d\tau$ reduces to the coordinate-time interval $dt$ and the resulting equations of motion for a free particle
\[
(\frac{d^2x^i}{dt^2})\left|\right._{\frac{v}{c} \to 0} = 0
\eqno{(1.32)}
\]
are only covariant with respect to  Galileo transformations.

\bigskip

\par\noindent
 Now, on the basis of Eq. (1.30), we assume that the gravitational field interacts with matter \underline{and} radiation through the general covariance which simply turns the ordinary derivative $\partial_\nu$ acting on any vector into the covariant derivative $D_\nu$ defined by
\[
(D_\nu)^\sigma_{\ \mu} \equiv \partial_\nu \delta^\sigma_{\ \mu} + \Gamma^\sigma_{\ \mu\nu}.
\eqno{(1.33)}
\]
In general, covariant derivatives do not commute in a curved space-time and we have
\[
[D_\mu , D_\nu]^\sigma_{\ \lambda} \equiv - R^\sigma_{\ \lambda\mu\nu} 
\eqno{(1.34)}
\]
where
\[
R^\sigma_{\ \lambda\mu\nu} \equiv \partial_\nu \Gamma^\sigma_{\ \lambda\mu} - \partial_\mu \Gamma^\sigma_{\ \lambda\nu} + \Gamma^\rho_{\ \lambda\mu} \Gamma^\sigma_{\ \rho\nu} - \Gamma^\rho_{\ \lambda\nu} \Gamma^\sigma_{\ \rho\mu}
\eqno{(1.35)}
\]
is the Riemann tensor. In the weak field approximation $|\frac{V}{c^2}| \ll 1$, the following space-time components of this curvature tensor
\[
R^i_{\ 00j} = - \frac{1}{c^2} \delta^{ik}  \partial_k \partial_j V(r) + \mathcal{O}(\frac{1}{c^4})
\eqno{(1.36)}
 \]
encode the first non-trivial gravitational effects of the Sun (and of the Moon) one   "feels"  on Earth, i.e., the tides: 
\[
V_{\textrm{{\tiny tide}}} (\vec x) \equiv \frac{1}{2} \sum_{k,j} x^k x^j \partial_k \partial_j V (\vec o).
\eqno{(1.37)}
\]
But what does  fix the full Riemann tensor in general:
\[
R^\sigma_{\ \lambda\mu\nu} \neq 0 \ ?
\eqno{(1.38)}
\]
Within a metric theory one can raise (lower) the space-time indices of any tensor. In particular, the anti-symmetry property of the Riemann tensor under a $\mu \leftrightarrow \nu$ interchange implies 
\[
D_\nu D_\mu R^{\sigma \ \mu\nu}_{\ \lambda} = 0 \ !
\eqno{(1.39)}
\]
These tensorial identities are most easily derived by working in a local inertial frame (i.e., $\Gamma\to 0$), as allowed by the Einstein equivalence principle.  It seems therefore quite interesting to focus our attention on the \underline{first} covariant derivatives of the Riemann tensor.

\bigskip

\par\noindent
In a conformally flat space-time, the metric $g_{\mu\nu} = A^2(V) \eta_{\mu\nu}$ only depends on a scalar gravitational field $V(t,\vec{x})$ and one easily derives the relation
\[
D_\mu R^{\sigma \ \mu\nu}_{\ \lambda} = \frac{1}{6} [\eta^{\sigma\nu} \eta_{\lambda \rho} - \delta^\nu_{\ \lambda} \delta^\sigma_{\ \rho}] \partial^\rho R
\eqno{(1.40)}
\]
with
\[
R \equiv g^{\lambda\mu} R^\nu_{\ \lambda\mu\nu} = -6 A^{-3} \  \square_\eta  \ A
\eqno{(1.41)}
\]
the curvature scalar. In our toy theory, i.e. the Nordstr\"om scalar theory based on Eq. (1.18), $A(V) = 1 + \frac{V}{c^2}$  and  massless gravitons freely propagate in a Minkowski fixed background. Consequently,  $R=0$ and  the Riemann tensor has to fulfil the non-trivial constraints
\[
D_\mu R^{\sigma \ \mu\nu}_{\ \lambda} = 0
\eqno{(1.42)}
\]
in the vacuum. Contrary to Eq.~(1.39), such non-linear constraints do \underline{not} result from the Einstein equi\-valence principle. We may thus conjecture that they are necessary to guarantee the strong version of the equivalence principle in any metric theory for gravitation \cite{8}.  

\par\noindent
It turns out that   Einstein's theory of gravity also complies with the tensorial constraints (1.42) in empty space. This property due to the purely geometrical Bianchi identities is quite remarkable since the gravi\-tational fields of general relativity are known to interact with themselves, even when propagating in the vacuum.  But in the presence of matter, what is then
\[
D_\mu R^{\sigma \ \mu\nu}_{\ \lambda} \equiv j^{\sigma \ \nu}_{\ \lambda}
\eqno{(1.43)}
\]
geometrically?  Well, astrophysics tells us that the Universe might be dominated by some dark matter at galactic distance scales and by some dark energy at cosmological distance scales. But these interpretations rely on the validity of general relativity at all scales, while direct evidences for such exotic substances are still missing. Consequently, alternative identifications of the $j^{\sigma \ \nu}_{\ \lambda}$ tensor are still allowed nowadays.

\bigskip

\par\noindent
In the Nordstr\"om scalar theory, we note from Eq. (1.40) that the conformally flat space-time background implies a genuine (mass) conservation law 
\[
\partial_\nu j^{\sigma \ \nu}_{\ \lambda} =  \frac{1}{6} [\eta^{\sigma\nu} \eta_{\lambda\rho} - \delta^\nu_{\ \lambda} \delta^\sigma_{\ \rho}] \partial_\nu \partial^\rho R = 0
\eqno{(1.44)}
\]
in a way analogous to the theory for electromagnetism. Indeed, the anti-symmetry property of the field strength in the inhomogeneous Maxwell equations
\[
\partial_\mu F^{\mu\nu} = j^\nu
\eqno{(1.45)}
\]
automatically implies the (charge) conservation law
\[
\partial_\nu j^\nu = 0,
\eqno{(1.46)}
\]
no matter the nature of the source at work (a Dirac electron, a Klein-Gordon charged pion,...).   However, a covariant conservation law like
\[
D_\nu j^{\sigma \ \nu}_{\ \lambda} = 0
\eqno{(1.47)}
\]
does not imply, in general, an exact differential conservation law \cite{9}. This is known to apply also for any non-abelian gauge theory to which we turn now.

\bigskip

\par\noindent
In \textbf{1954}, Yang and Mills examined what would happen if the isospin symmetry introduced to explain similarities of protons and neutrons were a local, i.e., space-time dependent, symmetry. For that purpose, they explored the possibility that the relative orientation of isospin at two distinct points of space-time has no physical meaning, (once of course electromagnetism is neglected). The local Lorentz frames of general relativity (labelled by Greek space-time indices) are thus simply replaced by local SU(2) frames (labelled by Latin internal indices) and a connection is needed to compare nucleons located at distinct points of space-time. In particular, the covariant derivative acting on any spinor $\Psi^b$ is introduced via the minimal substitution
\[
(D_\nu)^a_{\ b} \equiv \partial_\nu \delta^a_{\ b} - \ i\textrm{g} \ A^a_{\ b\nu}
\eqno{(1.48)}
\]

\medskip

\par\noindent
with $g$, the relevant coupling constant. To display the geometric nature of non-abelian gauge interactions, let us rescale the Yang-Mills  hermitian  matrix $A_\nu$ as follows:
\[
g \ A \to A.
\eqno{(1.49)}
\]
So, the components $\Gamma^\sigma_{\ \mu\nu}$ of the  connection are replaced by the massless gauge fields $A^a_{\ b\nu}$ and the Riemann-Christoffel curvature tensor $R^\sigma_{\ \lambda\mu\nu}$ by the non-abelian field strength $F^a_{\ b\mu\nu}$ such that:
\[
[D_\mu \ , \ D_\nu]^a_{\ b} \equiv -i \ F^a_{\ b\mu\nu}
\eqno{(1.50)}
\]
with
\[
F^a_{\ b\mu\nu} \equiv \partial_\mu A^a_{\ b\nu} - \partial_\nu A^a_{\ b\mu} +i A^c_{\ b\mu} A^a_{\ c\nu}  - i A^c_{\ b\nu} A^a_{\ c\mu}
\eqno{(1.51)}
\]

\medskip

\par\noindent
and $a, b,...$ isospin (or colours) indices. This parallel drawn between the space-time curvature in Eq. (1.35) and the non-abelian field strength in Eq.  (1.51) is quite striking. Note here that the appearance of a factor $i$ in the substitution 
\[
\Gamma \to -iA
\eqno{(1.52)}
\]
stems from the hermiticity of the $i\hbar\partial_\mu$ operator in quantum field theory.  The identities
\[
D_\nu \ D_\mu \ F^{a \ \mu\nu}_{\ b} = 0
\eqno{(1.53)}
\]
suggest that the universality of free-fall (i.e., the strong equivalence principle)  is on an equal footing with the universality of coupling (i.e., the gauge principle). To pursue such a parallel between gravitation and gauge interactions, we introduce external current densities 
\[
D_\mu \ F^{a \ \mu\nu}_{\ b} = j^{a \ \nu}_{\ b}.
\eqno{(1.54)}
\]
But again, what is $D_\mu \ F^{a \ \mu\nu}_{\ b}$ geometrically \cite{10}? Well, here   high-energy particle physics convincingly tells us that the gluons couple to  (spin $\frac{1}{2}$) matter fields, i.e, the coloured quarks. If we define the current as the first variation of the     Quantum Chromo-Dynamics (QCD) action with respect to the gauge fields, we obtain
\[
j^{a \ \nu}_{\ b} = \bar q_{b} \gamma^\nu q^a.
\eqno{(1.55)}
\]
It is then a direct consequence of the Dirac equation and its conjugate that this current indeed satisfies a  covariant conservation law given by
\[
D_\nu j^{a \ \nu}_{\ b} = \partial_\nu j^{a \ \nu}_{\ b} - i A^a_{\ c\mu} \ \   j^{c \ \mu}_{\ b} + i A^c_{\ b\mu} \ \   j^{a \ \mu}_{\ c} = 0.
\eqno{(1.56)}
\]
Yet, the current is not conserved in the ordinary sense because gauge fields carry the colours with which they interact.

\bigskip

\par\noindent
To summarize, the concepts of mass and energy in gravity provide us with to a deep connection between general coordinate transformations and gauge transformations, and in particular between general relati\-vity and non-abelian gauge theories. Einstein gravitational fields carry energy and thus gravitate the way Yang-Mills gauge fields carry colours and thus self-interact. This has to be contrasted with the Nordstr\"om massless graviton  which couples only to mass and the Maxwell neutral photon  which couples only to electric charge.

\subsection{Mass versus energy in electromagnetic, weak and strong interactions}

Today, Einstein's famous question
\begin{center}
\textit{Does the inertia of a body depend upon its energy content ?}
\end{center}
applies to all forms of  binding   energy $\Omega$  that contribute to the inertial mass $M$ of bound states: 
\[
M = \sum_i m_i + \frac{\Omega}{c^2}.
\eqno{(1.57)}
\] 
For compact spherical bodies of radius $R$, we already know from Eq. (1.8) that the gravitational   contribution to the binding energy per unit mass scales like $\frac{M}{R}$:
\[
\begin{array}{llllll}
s_{\textrm{{\tiny grav}}} \equiv |\frac{\Omega_{\textrm{{\tiny grav}}}}{Mc^2}| &\approx 10^{-26}&(\textrm{Sphere}..................M=2\times\!10^0 \ \textrm{kg}.......R=10~\textrm{cm})
\\
&\approx 10^{-10}&(\textrm{Earth}....................M=6\times\!10^{24}~\textrm{kg}......R=6400~\textrm{km})
\\
&\approx 10^{-6}&(\textrm{Sun}......................M=2\times\!10^{30}~\textrm{kg}......R=700000~\textrm{km})
\\
&\approx 10^{-3}&(\textrm{White Dwarf}.........M=2\times\!10^{30}~\textrm{kg}......R=1000~\textrm{km})
\\
&\approx 10^{-1}&(\textrm{Neutron Star}.........M=2\times\!10^{30}~\textrm{kg}......R=10~\textrm{km}).
\end{array}
\eqno{(1.58)}
\]
The ultimate stage of a heavy star, a stellar black hole, may thus be regarded as the extreme case where the binding energy is of the same order as the rest mass energy. For a dense stellar object with $R\approx \frac{2GM}{c^2}$, one indeed guesses  $s_{\textrm{{\tiny grav}}} \approx 0.3$ from Eq. (1.6).

\par\noindent
In the case of microscopic black holes, quantum arguments  plead in favour of a mass directly proportional to the Planck scale,
\[
M_{bh} \div (\frac{\hbar c}{G})^{\frac{1}{2}} \approx 10^{19}\UGeV{},
\eqno{(1.59)}
\]
excluding thus any production at the LHC if space-time is only 4-dimensional. For a binding energy proportional to the Newton constant $G$, the sensitivity can always be reexpressed as
\[
s_{\textrm{{\tiny grav}}}=-\frac{G}{M} \frac{\partial M}{\partial G}.
\eqno{(1.60)}
\]
Consequently, one derives now a firm upper bound for the ratio of internal gravitational binding energy to the total mass energy:
\[
s_{\textrm{{\tiny grav}}}  \leq \frac{1}{2},
\eqno{(1.61)}
\]
in full agreement with the field equations around a black hole for a tensor-scalar theory of gravity \cite{11}.  What about the other fundamental interactions?

\bigskip

\par\noindent
At the molecular level, mass defects in chemical reactions are known to be quite negligible since Lavoisier (\textbf{1789}):
\[
2 \textrm{H}_2+\textrm{O}_2 \to 2 \textrm{H}_2\textrm{O} + Q \ \ \ \ \textrm{with} \ \frac{Q}{Mc^2} \approx 10^{-13}.
\eqno{(1.62)}
\]
At the (sub) atomic level, such is not the case anymore. Electromagnetic, nuclear and strong interactions lead respectively to
\[
\begin{array}{lllll}\displaystyle
\textrm{-} \ \Bigl| \frac{\Omega_{\textrm{{\tiny em}}}}{Mc^2}\Bigr| \approx 10^{-8} 
& \displaystyle
 (m_{\textrm{{\tiny Hydrogen}}}  \approx m_{\textrm{{\tiny proton}}} + m_{\textrm{{\tiny electron}}} - 13.6\frac{\UeV{}}{c^2})
\\
&
\\ \displaystyle
\textrm{-} \  \Bigl|\frac{\Omega_{\textrm{{\tiny nucl}}}}{Mc^2}\Bigr| \approx 10^{-3} 
& \displaystyle
(m_{\textrm{{\tiny Deuterium}}}  \approx m_{\textrm{{\tiny proton}}} + m_{\textrm{{\tiny neutron}}} - 2.2\frac{\UMeV{}}{c^2})
\\
&
\\ \displaystyle
\textrm{-} \  \frac{E_{\textrm{{\tiny strong}}}}{Mc^2} \approx 1 
& \displaystyle
 (m_{\textrm{{\tiny proton}}} \approx m_{\textrm{{\tiny neutron}}} = + \  940\frac{\UMeV{}}{c^2}).
\end{array}
\eqno{(1.63)}
\]
As a result, the origin of the bulk of our mass and, consequently, of our weight is the kinetic energy of the massless gluons and nearly massless quarks confined in the nucleons. In technical words, our bathroom scales simply reacts to the fact that the QCD vacuum behaves like a paramagnetic medium \cite{12}.  The anti-screening effect of virtual gluons at $10^{-18}~m$ is also a superb answer of the strong interactions to Einstein's question about the inertia of a body:  gravitational self-interactions cannot saturate black holes mass the way strong interactions do for nucleons mass.

\section{Confinement and \mbox{\boldmath $[q_i,p_j] \neq 0$}}

\subsection{Nucleon mass}

In Section 1, we have seen that the strong interactions are based on a gauge invariant theory with Dirac particles (quarks) acting as colour sources:
\[
L_{\textrm{{\tiny fundamental}}} (\textrm{gluons; quarks}) = \bar{q} \ (i\gamma_\mu D^\mu -m) \ q.
\eqno{(2.1)}
\]
In the limit of two massless (up and down) quark flavours, chirality is conserved:
\[
\begin{array}{lllll}
q_{\textrm{{\tiny $L$}}} = \frac{1}{2}  (1-\gamma_5) q &,& \gamma_5 q_{\textrm{{\tiny $L$}}} = -q_{\textrm{{\tiny $L$}}}
\\
&&
\\
q_{\textrm{{\tiny $R$}}} =  \frac{1}{2}  (1+\gamma_5) q &,& \gamma_5 q_{\textrm{{\tiny $R$}}} = +q_{\textrm{{\tiny $R$}}}
\end{array}
\eqno{(2.2)}
\]
and the corresponding Lagrangian
\[
L_{\textrm{{\tiny fundamental}}} (\textrm{gluons}; \textrm{quarks}=u,d) = \bar{q}_{\textrm{{\tiny $L$}}} \ i\gamma_\mu D^\mu \ q_{\textrm{{\tiny $L$}}} + \bar{q}_{\textrm{{\tiny $R$}}} \ i\gamma_\mu D^\mu \ q_{\textrm{{\tiny $R$}}}
\eqno{(2.3)}
\]
is invariant under a global U(2)$_{\textrm{{\tiny $L$}}} \times$ U(2)$_{\textrm{{\tiny $R$}}}$ symmetry:
\[
\begin{array}{lllll}
q_{\textrm{{\tiny $L$}}} \to g_{\textrm{{\tiny $L$}}} \ q_{\textrm{{\tiny $L$}}} &\sim& (2_{\textrm{{\tiny $L$}}} , 1_{\textrm{{\tiny $R$}}})
\\
q_{\textrm{{\tiny $R$}}} \to g_{\textrm{{\tiny $R$}}} \ q_{\textrm{{\tiny $R$}}} &\sim& (1_{\textrm{{\tiny $L$}}} , 2_{\textrm{{\tiny $R$}}}).
\end{array}
\eqno{(2.4)}
\]
In Nature, one doublet of nucleon states $(J^{\textrm{{\tiny $P$}}}= \frac{1}{2}^+)$ turns out to be massive (1\UGeV{}) while one triplet of light pions $(J^{\textrm{{\tiny $P$}}} = 0^-)$ is observed around 100\UMeV{}. In other words, the chiral symmetry appears to be spontaneously broken down to an (approximate) SU(2) isospin symmetry through the confinement mechanism. In order to label the vacuum states, we introduce an effective (colour singlet) two-by-two  complex matrix $\chi$ which, by construction, transforms according to 
\[
\chi \to g_{\textrm{{\tiny $L$}}} \  \chi \  g_{\textrm{{\tiny $R$}}}^{\dag} \sim (2_{\textrm{{\tiny $L$}}} , 2^\ast_{\ \textrm{{\tiny $R$}}})
\eqno{(2.5)}
\]
with respect to the underlying chiral symmetry group. We may of course simply consider a bilinear in the up and down quark fields, 
\[
\chi^b_{\ a} \div \bar{q}_a (1-\gamma_5) q^b,
\eqno{(2.6)}
\]
though what really matters here are the chiral transformation properties. The complex field $\chi$   can always be expressed as a linear combination of two independent hermitian matrix fields $\sigma$ and $\pi$: 
\[
\chi \equiv \frac{(\sigma + i\pi)}{\sqrt{2}} \qquad (\sigma = \sigma^\alpha \tau_\alpha \ \ \ , \ \  \ \pi=\pi^\alpha \tau_\alpha)
\eqno{(2.7)}
\]
with $\tau_0$ the two-by-two unity matrix and $\tau_{1,2,3}$ the standard Pauli spin matrices: 
\[
\begin{pmatrix} 0&1\\1&0\end{pmatrix}, 
\begin{pmatrix} 0&-i\\i&0\end{pmatrix}, 
\begin{pmatrix} 1&0\\0&-1\end{pmatrix}.
\eqno{(2.8)}
\]
The effective Lagrangian for this field reads in general
\[
L_{\textrm{{\tiny effective}}} (\chi) =  \frac{1}{2}  \ \textrm{Tr} \ (\partial_\mu \chi \partial^\mu \chi^\dag) - V [\textrm{Tr} \ (\chi\chi^\dag)^n]
\eqno{(2.9)}
\] 
and the chiral invariant potential $V$ should provide $\chi$ with a non-zero real vacuum expectation value (v.e.v.) proportional to the unity matrix in order to preserve the isospin SU(2) subgroup characterized by $g_{\textrm{{\tiny $L$}}} = g_{\textrm{{\tiny $R$}}}$ vectorial transformations.

\bigskip

\par\noindent
For illustration,   we may consider a minimal linear sigma model
\[
L_{\textrm{{\tiny linear}}} (\chi) =  \frac{1}{2}   \ \textrm{Tr} \ (\partial_\mu \chi \partial^\mu \chi^\dag) -  \frac{\lambda}{4}   \ \textrm{Tr} \ (\chi\chi^\dag - \frac{f^2}{2})^2 \ \ , \lambda > 0
\eqno{(2.10)}
\]
where
\[
\begin{array}{llll}
< 0 | \sigma | 0 > \ = f 1\hspace{-0.7mm}\raisebox{0.5mm}{$\scriptstyle |$}
\\
< 0 | \pi | 0 > \ = 0.
\end{array}
\eqno{(2.11)}
\]
A suitable redefinition of the $\sigma$ field,
\[
\sigma \to \sigma - <0|\sigma|0>, 
\eqno{(2.12)}
\]
leads then to the following physical mass spectrum
\[
\begin{array}{llll}
m_{\sigma^\alpha} = \sqrt{\lambda} f
\\
m_{\pi^\alpha} = 0.
\end{array}
\eqno{(2.13)}
\]
\underline{If} in addition, we assume that the nucleon doublet 
\[
N = \begin{pmatrix} p\\ \\ n\end{pmatrix}
\eqno{(2.14)}
\]
transforms as
\[
\begin{array}{llll}
N_{\textrm{{\tiny $L$}}} \to g_{\textrm{{\tiny $L$}}} N_{\textrm{{\tiny $L$}}} &\sim& (2_{\textrm{{\tiny $L$}}} , 1_{\textrm{{\tiny $R$}}})
\\
N_{\textrm{{\tiny $R$}}} \to g_{\textrm{{\tiny $R$}}} N_{\textrm{{\tiny $R$}}} &\sim& (1_{\textrm{{\tiny $L$}}} , 2_{\textrm{{\tiny $R$}}})
\end{array}
\eqno{(2.15)}
\]
under the chiral symmetry group, we may also consider
\[
\begin{array}{llll}
L_{\textrm{{\tiny linear}}} (N) &=& \overline{N}_{\textrm{{\tiny $L$}}} \ i\gamma_\mu \partial^\mu \ N_{\textrm{{\tiny $L$}}} + \overline{N}_{\textrm{{\tiny $R$}}} \ i\gamma_\mu\partial^\mu N_{\textrm{{\tiny $R$}}} - g_{\textrm{{\tiny $\pi$NN}}} (\overline{N}_{\textrm{{\tiny $L$}}} \chi N_{\textrm{{\tiny $R$}}} + \textrm{h.c.})
\\
&&
\\
&=&\displaystyle \overline{N} \ i\gamma_\mu\partial^\mu N - \frac{g_{\textrm{{\tiny $\pi$NN}}}}{\sqrt{2}} (\overline{N} \sigma N + i \overline{N} \gamma_5 \pi N)
\end{array}
\eqno{(2.16)}
\]
with $\frac{g_{\textrm{{\tiny $\pi$NN}}}}{\sqrt{2}} \approx 13.5$, the measured pseudoscalar coupling. The $\sigma$ and $\pi$ are then identified as scalar $(0^+)$ and pseudoscalar $(0^-)$ fields, respectively, while the nucleon mass is driven by the v.e.v. of the $\sigma$ field given in Eq. (2.11) to fulfil the relation
\[
M_{\textrm{{\tiny $N$}}} = g_{\textrm{{\tiny $\pi$NN}}}  \frac{f}{\sqrt{2}}.
\eqno{(2.17)}
\]
The rather simple linear sigma model defined by (2.10) and (2.16) seems to correctly implement the chiral symmetry breaking since it produces a (semi) realistic mass spectrum for the  pseudoscalar triplet $\pi$,   the nucleon doublet $N$ and the scalar triplet $a_0$:
\[
140\UMeV{} = m_\pi \ll M_N \approx m_{a_0} = 980\UMeV{}.
\eqno{(2.18)}
\]
However, at the experimental level, the full scalar multiplet around the nucleon mass scale is not settled yet. Moreover, at the theoretical level, chiral transformations of baryons are ambiguous. This latter fact becomes particularly obvious in the generalized case of three massless quark flavours (u, d, s). The Gell-Mann baryon octet $(J^{\textrm{{\tiny $P$}}}= \frac{1}{2}^+)$ 
\[
B =
\begin{pmatrix} \Sigma^0+\frac{\Lambda}{\sqrt{3}}   &    \sqrt{2}\Sigma^+   &   \sqrt{2}p
\\
\sqrt{2}\Sigma^-   &  -\Sigma^0+\frac{\Lambda}{\sqrt{3}}   &  \sqrt{2}n
\\
\sqrt{2}\Xi^-   &  \sqrt{2} \Xi^0   & \frac{-2\Lambda}{\sqrt{3}}
\end{pmatrix}
\eqno{(2.19)}
\]
may indeed transform either as
\[
\begin{array}{llll}
B_{\textrm{{\tiny $L$}}} \to g_{\textrm{{\tiny $L$}}} B_{\textrm{{\tiny $L$}}} g_{\textrm{{\tiny $L$}}}^{\dag} &\sim& (8_{\textrm{{\tiny $L$}}} , 1_{\textrm{{\tiny $R$}}})
\\
B_{\textrm{{\tiny $R$}}} \to g_{\textrm{{\tiny $R$}}} B_{\textrm{{\tiny $R$}}} g_{\textrm{{\tiny $R$}}}^{\dag} &\sim& (1_{\textrm{{\tiny $L$}}} , 8_{\textrm{{\tiny $R$}}})
\end{array}
\eqno{(2.20)}
\]
or as
\[
\begin{array}{llll}
B_{\textrm{{\tiny $L$}}} \to g_{\textrm{{\tiny $L$}}} B_{\textrm{{\tiny $L$}}} g_{\textrm{{\tiny $R$}}}^{\dag} &\sim& (3_{\textrm{{\tiny $L$}}} , 3^\ast_{\textrm{{\tiny $R$}}})
\\
B_{\textrm{{\tiny $R$}}} \to g_{\textrm{{\tiny $R$}}} B_{\textrm{{\tiny $R$}}} g_{\textrm{{\tiny $L$}}}^{\dag} &\sim& (3^\ast_{\textrm{{\tiny $L$}}} , 3_{\textrm{{\tiny $R$}}})
\end{array}
\eqno{(2.21)}
\]
under SU(3)$_{\textrm{{\tiny $L$}}} \times$ SU(3)$_{\textrm{{\tiny $R$}}}$ since only the transformation properties of the baryon under the vectorial subgroup SU($n_{\textrm{{\tiny $F$}}}$) (isospin symmetry, eightfold way,...) really matter \cite{13}. So, let us turn to a non-linear effective theory to get rid of the elusive scalars and couple baryons to pseudoscalars in a unique way. For that purpose, we make use of the polar theorem which tells us that any arbitrary matrix $(\xi^\dag\chi)$ can be written as the product of a hermitian matrix $(\sigma)$ and a unitary matrix $(\xi)$:
\[
\chi \equiv \xi (\pi) \frac{\sigma}{\sqrt{2}} \xi (\pi).
\eqno{(2.22)}
\]
The main advantage of this new parametrization is that now the most general potential only depends on the scalar fields:
\[
V[\textrm{Tr} \ (\chi\chi^\dag)^n] = V(\sigma).
\eqno{(2.23)}
\]
The chiral transformations (2.5) of $\chi$ require
\[
\xi \to g_{\textrm{{\tiny $L$}}} \xi h^\dag = h \  \xi g^{\dag}_{\textrm{{\tiny $R$}}}
\eqno{(2.24)}
\]
such that these scalars transform linearly with respect to $h$:
\[
\sigma \to h(x) \sigma \ h(x)^\dag.
\eqno{(2.25)}
\]
The vectorial transformations $h$ are not broken by the v.e.v. of the $\sigma$ field given in Eq. (2.11). Moreover, being non-linear functions of $g_{\textrm{{\tiny $L$}}}$, $g_{\textrm{{\tiny $R$}}}$ \underline{and} $\pi(x)$, they depend in general on the space-time coordinates. Yet, for $g_{\textrm{{\tiny $L$}}} = g_{\textrm{{\tiny $R$}}}$, we have $h = g_{\textrm{{\tiny $L$}}} = g_{\textrm{{\tiny $R$}}}$ and we recover the successful SU(2)$_{\textrm{{\tiny $I$}}} \times$ U(1)$_{\textrm{{\tiny $B$}}}$  global symmetry. It is thus quite natural to extend these linear, though local, transformations to all the other hadron isospin multiplets to describe their interactions with the light pseudoscalar one. In particular, we shall impose this local   \textit{hidden symmetry} on the nucleons: 
\[
N \to h(x) \ N.
\eqno{(2.26)}
\]
From the following transformation laws
\[
\begin{array}{llll}
(\xi^\dag \partial_\mu \xi) \to h(\xi^\dag \partial_\mu \xi) h^\dag + h (\partial_\mu) h^\dag
\\
(\xi \partial_\mu \xi^\dag) \to h(\xi \partial_\mu \xi^\dag) h^\dag + h(\partial_\mu) h^\dag,
\end{array}
\eqno{(2.27)}
\]
one can indeed easily build a gauge invariant effective Lagrangian for the nucleon-pion interactions. At the leading order in the derivative couplings, it reads
\[
L_{\textrm{{\tiny non-linear}}} (\overline{N},\pi) = \overline{N} (i\ \gamma^\mu D_\mu - M_{\textrm{{\tiny $N$}}}) N + g_{\textrm{{\tiny $A$}}} \  \overline{N} \ \gamma^\mu\gamma_5 \ A_\mu \ N
\eqno{(2.28)}
\]
with
\[
D_\mu \ N \equiv [\partial_\mu + \frac{1}{2}  (\xi^\dag \partial_\mu \xi + \xi \partial_\mu \xi^\dag)] N \to h(x) D_\mu \ N
\eqno{(2.29)}
\]
the effective covariant derivative acting on the nucleon doublet and
\[
A_\mu \equiv  \frac{i}{2}  (\xi^\dag \partial_\mu \xi - \xi \partial_\mu \xi^\dag) \to h A_\mu h^\dag
\eqno{(2.30)}
\]
an effective field coupled to the axial-vector nucleon current.

\bigskip

\par\noindent
If the elusive scalar degrees of freedom are frozen at their v.e.v., they simply decouple and we are left with an effective theory for the light pseudoscalar fields alone:
\[
L_{\textrm{{\tiny non-linear}}} (\pi) = -f^2 \ \textrm{Tr} \ (A_\mu A^\mu) = \frac{f^2}{4} \ \textrm{Tr} \ (\partial_\mu   U \ \partial^\mu   U^\dag)
\eqno{(2.31)}
\]
with 
\[
U \equiv \xi^2 \to g_{\textrm{{\tiny $L$}}} \ U \ g^{\dag}_{\textrm{{\tiny $R$}}}.
\eqno{(2.32)}
\]
This minimal effective Lagrangian contains in fact all the necessary features of the spontaneous chiral symmetry breaking pattern
\[
\textrm{U(2)}_{\textrm{{\tiny $L$}}} \times \textrm{U(2)}_{\textrm{{\tiny $R$}}} \to \textrm{SU(2)}_{\textrm{{\tiny isospin}}} \times \textrm{U(1)}_{\textrm{{\tiny baryon}}}.
\eqno{(2.33)}
\]
Indeed, if we expand the $U$ field as follows:  
\[
U(\pi) = 1\hspace{-0.7mm}\raisebox{0.5mm}{$\scriptstyle |$}+i(\frac{\pi}{f}) -  \frac{1}{2}  (\frac{\pi}{f})^2 - i \ a(\frac{\pi}{f})^3 + (a-\frac{1}{8})(\frac{\pi}{f})^4...
\eqno{(2.34)}
\]

\begin{enumerate}
\item[-]    the vacuum expectation value of $U$ is invariant under the unbroken vectorial subgroup U(2)$_{\textrm{{\tiny $L+R$}}}$ defined by $g_{\textrm{{\tiny $L$}}} = g_{\textrm{{\tiny $R$}}}$:
\[
g_{\textrm{{\tiny $L$}}} < 0|U|0 > g_{\textrm{{\tiny $R$}}}^{\dag} = \  < 0|U|0 >;
\eqno{(2.35)}
\]
\item[-] the excited fields out of the vacuum are the four pseudoscalar Goldstone bosons associated with the four broken generators:
\[
\pi=
\begin{pmatrix} \pi_3+\eta_0&\sqrt{2}\pi^+
\\
\sqrt{2} \pi^-&-\pi_3+\eta_0\end{pmatrix}.
\eqno{(2.36)}
\]
\end{enumerate}

\par\noindent
The unitarity condition
\[
UU^\dag=1\hspace{-0.7mm}\raisebox{0.5mm}{$\scriptstyle |$}
\eqno{(2.37)}
\]
is fulfilled for any real value of $a$. It is quite convenient (and also standard) to fix the value of this free parameter to 
\[
a=\frac{1}{6}
\eqno{(2.38)}
\]
with
\[
U(\pi) = \exp (\frac{i\pi}{f}).
\eqno{(2.39)}
\]
But you may as well choose for example the value
\[
a = \frac{1}{4}
\eqno{(2.40)}
\]
with
\[
U (\pi)=\frac{(1+\frac{i\pi}{2f})}{(1-\frac{i\pi}{2f})}
\eqno{(2.41)}
\]
since chiral invariance ensures that any physical quantity is $a$-independent.

\bigskip

\par\noindent
Expanding $A_\mu$ defined in Eq. (2.30) to first order in $\pi$, we note that the derivative nucleon-pion interaction is related to the standard pseudoscalar one through the Dirac equation of motion and implies the  Goldberger-Treiman relation \cite{14}
\[
\frac{g_{\textrm{{\tiny $\pi$NN}}}}{\sqrt{2}} = g_{\textrm{{\tiny $A$}}} \frac{M_{\textrm{{\tiny $N$}}}}{f}.
\eqno{(2.42)}
\]
From $g_{\textrm{{\tiny $A$}}} \approx 1.27$, the axial-vector coupling measured in parity-violating $(n\to pe \bar \nu)$  $\beta$ decays, and $M_{\textrm{{\tiny $N$}}} \approx 940\UMeV{}$, the average mass of the nucleons, one can already infer that $f \approx 90\UMeV{}$ for the v.e.v. of the $\sigma$ field defined in Eq. (2.11). However, a more precise estimate of the remaining free parameter $f$ is directly obtained from weak interactions. Indeed, gauging  $\textrm{SU(2)}_{\textrm{{\tiny $L$}}}$ requires, as usual, the introduction of a covariant derivative. At the fundamental level, it amounts to the minimal substitution 
\[
D_\mu \to D_\mu -i W^{\textrm{{\tiny $L$}}}_{\ \ \mu}
\eqno{(2.43)}
\]
for the left-handed component of the quark fields in Eq. (2.1), such that

\[
L_{\textrm{{\tiny fund.}}} (q) \ni \bar{q}_{\textrm{{\tiny $L$}}}^{\ a} \gamma^\mu W^{\textrm{{\tiny $L$}}\ \ ab}_{\ \ \mu} q_{\textrm{{\tiny $L$}}}^{\ b} \equiv J_{\ \ \textrm{{\tiny $L$}}}^{\mu} (q) W_{\ \ \mu}^{\textrm{{\tiny $L$}}}
\eqno{(2.44)}
\]
with
\[
(J_{\ \ \textrm{{\tiny $L$}}}^{\mu})^{\textrm{{\tiny $ba$}}} (q) = \bar{q}_{\textrm{{\tiny $L$}}}^{\ a} \gamma^\mu q_{\textrm{{\tiny $L$}}}^{\ b}.
\eqno{(2.45)}
\]

\par\noindent
At the effective level, we have to consider the minimal substitution 
\[
\partial_\mu U \to D_\mu U = \partial_\mu U - i W^{\textrm{{\tiny $L$}}}_{\ \ \mu} U
\eqno{(2.46)}
\]
in Eq. (2.31) since
\[
U \to g_{\textrm{{\tiny $L$}}} (x) U
\eqno{(2.47)}
\]
under SU(2)$_{\textrm{{\tiny $L$}}}$ gauge transformations. The interaction terms are given by
\[
L_{\textrm{{\tiny eff.}}} (\pi) \ni  -i  \frac{f^2}{4}   \ \textrm{Tr} \ (W^{\textrm{{\tiny $L$}}}_{\ \ \mu} \ U \partial^\mu U^\dag - \partial^\mu UU^\dag W^{\textrm{{\tiny $L$}}}_{\ \ \mu}) \equiv J^\mu_{\ \textrm{{\tiny $L$}}} (\pi) W^{\textrm{{\tiny $L$}}}_{\ \mu}
\eqno{(2.48)}
\]
with
\[
(J^\mu_{\ \ \textrm{{\tiny $L$}}})^{\textrm{{\tiny $ba$}}} (\pi) = i  \frac{f^2}{2}  (\partial^\mu UU^\dag)^{\textrm{{\tiny $ba$}}} \ni - \frac{f}{2}  \partial^\mu \pi^{ba}
\eqno{(2.49)}
\]

\par\noindent
the left-handed hadronic current. Consequently,  we obtain the vacuum-to-pion hadronic matrix element
\[
< 0 | (J^\mu_{\ \ \textrm{{\tiny $L$}}})^{ud} | \pi^+ > \  =-i  \frac{f}{\sqrt{2}}  p^\mu
\eqno{(2.50)}
\]
with
\[
f = f_\pi \approx 93\UMeV{}
\eqno{(2.51)}
\]

\par\noindent
extracted from the measured    $\pi^+ \to e^+ \nu_{\textrm{{\tiny $e$}}}$ decay amplitude.

\bigskip

\par\noindent
So, now we dispose of a rather elegant and very efficient frame to incorporate all the well-known results originally obtained from standard current algebra techniques, and in particular the theorems on electromagnetic quantum corrections derived in the sixties.  Electromagnetism is indeed introduced through the minimal substitution
\[
\partial_\mu U \to D_\mu U = \partial_\mu U -i V_\mu [Q,U], \qquad Q = e \ \textrm{diag} \ (+\frac{2}{3} , -\frac{1}{3})
\eqno{(2.52)}
\]
since
\[
U \to g(x) \ U g(x)^\dag
\eqno{(2.53)}
\]

\par\noindent
under vectorial U(1)$_{\textrm{{\tiny $QED$}}}$ gauge transformations. In the  Landau gauge for the photon propagator, the only relevant one-loop diagram is a quadratically divergent tadpole produced by the contact term  
\[
L(U) \ni   \frac{e^2f^2}{2} \ \textrm{Tr} \ (QUQU^\dag) V_\mu \ V^\mu,
\eqno{(2.54)}
\]
with $U(\pi)$ defined in Eq. (2.34).  Consequently,
\begin{enumerate}
\item[-]   expanding the $U$ field at $\mathcal{O}(\pi^2)$, we obtain the combination Tr$(QQ\pi^2 - Q\pi Q\pi)$ which implies a mass correction for the charged pion only:
\[
m^2_{\pi^+} - m^2_{\pi^0} =  \frac{3\alpha}{4\pi}   \Lambda^2  \ \ ;
\eqno{(2.55)}
\]

\item[-] expanding the U field at $\mathcal{O}(\pi^4)$, with $a = 0$ to get rid of the cubic term, we obtain the combination Tr$(QQ\pi^4 - Q\pi^2Q\pi^2)$ which does not allow an iso-singlet to decay into three pions, i.e.,
\[
A^{\textrm{{\tiny e.m.}}} (\eta_0 \to \pi^+\pi^0\pi^-)=0.
\eqno{(2.56)}
\]
\end{enumerate}

\medskip

\par\noindent
The knowledge of the underlying QCD theory helped us understanding these two  puzzling results. On the one hand, the quadratic dependence on the ultraviolet momentum cut-off $\Lambda$ in the $\pi^+ -\pi^0$ mass difference is tamed in a natural way by the vector and axial vector resonances at work around 0.8 \UGeV{}. (Interestingly, the composite structure of the pion softens its electromagnetic self-energy the way a composite structure for the Higgs scalar would naturally protect its mass in an effective theory of electroweak interactions). On the other hand, the observed isospin-violating $\eta^{(')} \to \pi\pi\pi$  decays are induced by the up-down quark mass difference to which we turn now.

\subsection{Nucleon mass splitting}

At the fundamental level, isospin violation beyond electromagnetism arises from the mass term
\[
\Delta L_m (u,d) = - \bar{q}_{\textrm{{\tiny $L$}}a} \  m^a_{\ b} \  q_{\textrm{{\tiny $R$}}}^{\ b} + \textrm{h.c.}
\eqno{(2.57)}
\]
If the two-by-two quark mass matrix $m$ is first treated as a spurion field, it   has  to transform under the chiral U(2)$_{\textrm{{\tiny $L$}}}$ $\times$ U(2)$_{\textrm{{\tiny $R$}}}$ group according to the rule
\[
m(x) \to g_{\textrm{{\tiny $L$}}} \  m(x) g_{\textrm{{\tiny $R$}}}^{\dag}.
\eqno{(2.58)}
\]
At the effective level, the leading mass correction for the nucleons arises from the chiral invariant 
\[
\Delta L_m (N) = - \frac{b}{2}  \overline{N}(\xi^\dag m\xi^\dag + \xi m^\dag \xi)N  \ni  - \frac{b}{2}   \overline{N}(m+m^\dag)N.
\eqno{(2.59)}
\]
Once the quark mass matrix is frozen to its real eigenvalues
\[
m = \begin{pmatrix}
m_u & 0 \\ 0 & m_d
\end{pmatrix}
\eqno{(2.60)}
\]
a neutron-proton splitting takes then place with
\[
m_n - m_p = b (m_d - m_u) \approx 1.3 \UMeV{}
\eqno{(2.61)}
\]
if electromagnetic self-interaction corrections (in principle favourable to the proton) are neglected.  Correspondingly, the leading mass correction for the (pseudo-) Goldstone bosons arises from
\[
\Delta L_m (\pi) =  \frac{f^2r}{4}  \ \textrm{Tr} \ (mU^\dag + Um^\dag) \ni -  \frac{r}{4}  \ \textrm{Tr} \ (m\pi^2).
\eqno{(2.62)}
\]
For the charged pions, we obtain
\[
m_{\pi^\pm}^{2} =  \frac{r}{2}  (m_u + m_d) \approx 140 \UMeV{}.
\eqno{(2.63)}
\]
From the trace of the neutral pseudoscalars squared mass matrix 
\[
m_{\textrm{\ {\tiny neutral}}}^2 =  \frac{r}{2}  
\begin{pmatrix}
m_u + m_d & m_u - m_d
\\
&
\\
m_u - m_d & m_u + m_d
\end{pmatrix},
\eqno{(2.64)}
\]
we also obtain a quadratic mass relation
\[
m_{\pi^0}^{2} + m_{\eta^\prime}^{2} = 2m_{\pi^\pm}^{2} 
\eqno{(2.65)}
\]
in clear contradiction with the observed mass spectrum
\[
\begin{array}{lll}
m_{\pi^0}  &=& 135 \UMeV{}   \\
m_{\eta^\prime}  &=& 958 \UMeV{}.   
\end{array}
\eqno{(2.66)}
\]
The fact that the $\eta^\prime$  mass is close to the nucleon (and scalar) mass scale given in Eq. (2.18) strongly suggests the way to solve this problem  \cite{15}: assume the symmetry breaking pattern to be
\[
\textrm{SU(2)}_{\textrm{{\tiny $L$}}} \times 
\textrm{SU(2)}_{\textrm{{\tiny $R$}}} \times
\textrm{U(1)}_{\textrm{{\tiny $B$}}}
\to
\textrm{SU(2)}_{\textrm{{\tiny $I$}}} \times
\textrm{U(1)}_{\textrm{{\tiny $B$}}}
\eqno{(2.67)}
\]
instead of (2.33),  such that only three Goldstone bosons are produced and not four! To implement the explicit breaking of the axial U(1), we thus add an $\mathcal{O}$(1\UGeV{}) mass term for the iso-singlet $\eta_0$:
\[
\Delta L_{\textrm{{\tiny U(1)}}} =  -  \frac{1}{2}  m_0^{2} \  \eta_0^{\ 2}. 
\eqno{(2.68)}
\]
The squared mass matrix becomes then
\[
m_{\textrm{\ {\tiny neutral}}}^2 =  
\begin{pmatrix}
m_{\pi^\pm}^{2}  & \Delta
\\
&
\\
\Delta & m_0^{2} + m_{\pi^\pm}^{2}\end{pmatrix} ;  |\Delta| \equiv  \frac{r}{2} | m_u - m_d | \ll  m_0^{2}
\eqno{(2.69)}
\]
and the resulting quadratic mass relations
\[
\begin{array}{llll}
& m_{\pi^0}^{2} \approx 
m_{\pi^\pm}^{2} - \displaystyle \frac{\Delta^2}{m_{0}^{\ 2}}
\\
&
\\
& m_{\eta^\prime}^{2} \approx m_{0}^{2} + m_{\pi^\pm}^{2} + \displaystyle \frac{\Delta^2}{m_{0}^{2}}
\end{array}
\eqno{(2.70)}
\]
are in agreement with the electromagnetic self-interaction correction given in Eq. (2.55). But we still have to check that the modification (2.68) of the effective theory for strong interactions is compatible with what we know from the underlying QCD dynamics. Let us for that purpose consider the conservation law of the iso-singlet current
\[
J^\mu_{\ 5} \equiv   \bar{u} \gamma^\mu \gamma_5 u + \bar{d} \gamma^\mu \gamma_5 d
\eqno{(2.71)}
\]
associated with the axial U(1) symmetry. At the effective level, the right-handed current $J^\mu_{\ \textrm{{\tiny $R$}}}$ is directly obtained from the simple parity transformation
\[
U(\pi) \stackrel{P}{\longrightarrow} U^\dag(\pi) = U(-\pi)
\eqno{(2.72)}
\]
applied on the left-handed hadronic current $J^\mu_{\ \textrm{{\tiny $L$}}}$   already derived in Eq. (2.49), such that
\[
J^\mu_{\ 5} \equiv \ \textrm{Tr}\  (J^\mu_{\ \textrm{{\tiny $R$}}} 
- J^\mu_{\ \textrm{{\tiny $L$}}}) = i  \frac{f^2}{2} \ \textrm{Tr}\  (\partial^\mu U^\dag U - \partial^\mu UU^\dag).
\eqno{(2.73)}
\]
From the identity
\[
\textrm{Tr}\  (\partial_\mu UU^+) =  \frac{2i}{f} \partial_\mu \eta_0,
\eqno{(2.74)}
\]
we infer that the iso-singlet current is \underline{not} conserved in the massless limit $m_u = m_d = 0$:
\[
\partial_\mu J^\mu_{\ 5} = 2 f \ \square \   \eta_0 = -2 f m_0^{2} \eta_0.
\eqno{(2.75)}
\]
At the fundamental level, the same violation of a classical conservation law is induced by quantum effects and the so-called axial $U(1)$ anomaly is precisely given by
\[
\partial_\mu J^\mu_{\ 5} = n_{\textrm{{\tiny $F$}}}  \frac{\alpha_s}{4\pi}  G^{\alpha\beta}  \widetilde{G}_{\alpha\beta} 
\eqno{(2.76)}
\]
with
\[
\widetilde{G}_{\alpha\beta} \equiv  \frac{1}{2} \varepsilon_{\alpha\beta\gamma\delta} G^{\gamma\delta}
\eqno{(2.77)}
\]
the dual of the gluon field strength. But this is not the end of the story since, as we shall see, the Standard Model for electroweak interactions provides us in principle with a \underline{complex} quark mass matrix
\[
m  \neq m^\dag
\eqno{(2.78)}
\]
via the Higgs mechanism. Up to now, specific chiral $g^0_{\ \textrm{{\tiny $L$,$R$}}}$ unitary transformations had been implicitly used to write this mass matrix as a diagonal \underline{and} real one. But the axial U(1) anomaly implies that one phase cannot be rotated away and that we end up at best with
\[
g^0_{\ \textrm{{\tiny $L$}}} m g^{0 \ \ \dag}_{\ \textrm{{\tiny $R$}}} =
\exp (\frac{i\theta_{\textrm{{\tiny $M$}}}}{4})
\begin{pmatrix}
m_u & 0\\
0 & m_d
\end{pmatrix}
\exp (\frac{i\theta_{\textrm{{\tiny $M$}}}}{4}).
\eqno{(2.79)}
\]
The presence of a physical phase is in principle the signal for a violation under time-reversal. The corres\-ponding $T$ operator is indeed   anti-unitary, as most easily seen from its effect on the Heisenberg commutator
\[
[q_i,p_j] = i\hbar \ \  \delta_{ij} \stackrel{{\textrm{{\tiny $T$}}}}{\to} -i\hbar \ \delta_{ij} = [q_i, -p_j].
\eqno{(2.80)}
\]
Note that this microscopic irreversibility has to be distinguished from macroscopic ones which originate in quite peculiar boundary conditions: a Bunsen burner for heat propagation or the Lema\^\i tre Big Bang for an expanding Universe...

\bigskip

\par\noindent
 A  simple way to convince ourselves that the strong axial anomaly indeed implies $T$-violation is through the field redefinition
\[
U \to \exp (\frac{i\theta_{\textrm{{\tiny $M$}}}}{2}) U
\eqno{(2.81)}
\]
or, equivalently,
\[
\eta_0 \to \eta_0 +  \frac{f}{2}  \theta_{\textrm{{\tiny $M$}}}
\eqno{(2.82)}
\]
with
\[
\theta_{\textrm{{\tiny $M$}}} \equiv \arg \det m.
\eqno{(2.83)}
\]
This field redefinition renders $m$ totally real in Eq. (2.62) but modifies of course the anomalous part (2.68) of the effective Lagrangian,
\[
\Delta L_{\textrm{{\tiny U(1)}}} \to \Delta L_{\textrm{{\tiny U(1)}}} -  \frac{f}{2}  m_0^{2} \theta_{\textrm{{\tiny $M$}}} \eta_0,
\eqno{(2.84)}
\]
in such a way that the $\eta_0$ pseudoscalar field gets now a non-zero v.e.v.:
\[
<0 | \eta_0 (0^{-+}) | 0 > \approx - [\frac{m_0^{2}}{m_0^{2} + m_\pi^{2}}] \frac{f\theta_{\textrm{{\tiny $M$}}}}{2}.
\eqno{(2.85)}
\]
Accordingly, both $T$ \underline{and} $P$-violations occur in strong interactions once $m_0^{2} \neq 0$. The identification of the axial anomaly,  expressed at the effective level in Eq. (2.75) and at the fundamental one in Eq. (2.76), together with the shift in Eq. (2.84) require a  corresponding modification of the QCD action itself:
\[
L_{\textrm{{\tiny QCD}}} \to L_{\textrm{{\tiny QCD}}} + \frac{\alpha_s}{8\pi}  \theta_{\textrm{{\tiny $M$}}} G^{\alpha\beta} \widetilde{G}_{\alpha\beta}.
\eqno{(2.86)}
\]
So, this new pseudoscalar term implies physical effects despite the fact that $G\widetilde{G}$ can be written as a total derivative. Let us illustrate this rather surprising result with a first example.

\bigskip

\par\noindent
If we choose $a = \frac{1}{6}$ in Eq. (2.34), we may ignore the kinetic term in Eq. (2.31) and focus on the mass term in Eq. (2.62), with
\[
\Delta L_m (\pi) \ni  \frac{m^{2}_\pi}{2f^2} \  \eta_0 \pi^+ \pi^- \eta_0
\eqno{(2.87)}
\]
in the isospin limit, to get a non-zero $T$-violating $\eta' \to \pi^+\pi^-$ decay amplitude. Let indeed one of the two $\eta_0$'s propagate and then annihilate into the vacuum via the linear term in $\eta_0$ introduced in Eq. (2.84):

\vspace*{-5mm}

\begin{center}
\includegraphics[]{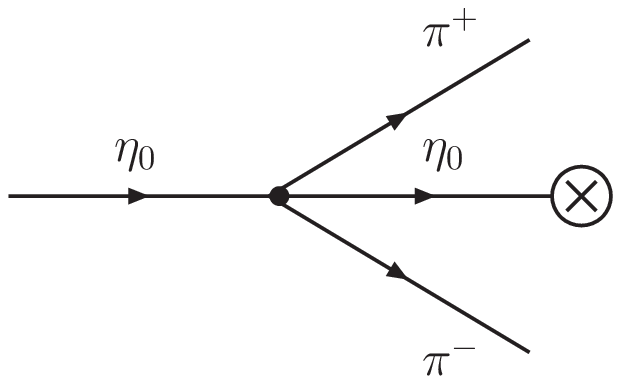}
\end{center}

\par\noindent
This non-local "tadpole" contribution amounts to substitute \underline{directly} $<0|\eta_0|0>$ for one $\eta_0$ in Eq. (2.87) and we obtain in that manner the local amplitude  
\[
| A (\eta' \to \pi^+\pi^-)| =  \frac{m_\pi^{2}}{2f}  [\frac{m_0^{2}}{m_0^{2} + m_\pi^{2}}] \theta_{\textrm{{\tiny $M$}}}.
\eqno{(2.88)}
\]
The corresponding two-body decay width reads
\[
\Gamma (\eta' \to \pi^+\pi^-) \equiv \frac{1}{16\pi m_{\eta'}}  |
A(\eta' \to \pi^+\pi^-) |^2  [1-4 \frac{m_\pi^{2}}{m_{\eta'}^{2}}]^{\frac{1}{2}} \approx 0.2\times\!\theta^2_{\textrm{{\tiny $M$}}} \UMeV{}.
\eqno{(2.89)}
\]
Taking into account the measured $\eta'$  total width, we obtain 
\[
\textrm{Br} \ (\eta' \to \pi^+\pi^-) \approx \theta^2_{\textrm{{\tiny $M$}}}
\eqno{(2.90)}
\]
such that the present experimental limit on this branching ratio
\[
\textrm{Br} \ (\eta' \to \pi^+\pi^-) < 2\times\!10^{-2}
\eqno{(2.91)}
\]
provides a rather weak bound
\[
\theta_{\textrm{{\tiny $M$}}} \lesssim 10^{-1}.
\eqno{(2.92)}
\]
Note however that the sizeable $\eta_0$ component in $\eta$(548) extracted from the non-linear effective theory with three light quark flavours $(u, d, s)$ \cite{16}, 
\[
\begin{array}{lllll}
\eta = \eta_8 \cos \phi - \eta_0 \sin \phi \\
\qquad \qquad \qquad \qquad \qquad \qquad \qquad \qquad \qquad (\phi \approx -22^\circ),\\
\eta' = \eta_8 \sin \phi + \eta_0 \cos \phi
\end{array}
\eqno{(2.93)}
\]
allows us to get a stronger bound, namely
\[
\theta_{\textrm{{\tiny $M$}}} < 3\times\!10^{-4},
\eqno{(2.94)}
\]
from the new experimental limit
\[
\textrm{Br} \ (\eta \to \pi^+\pi^-) <  1.3\times\!10^{-5}.
\eqno{(2.95)}
\]
Significant improvements on these tree-level bounds are not foreseen since branching ratios are \underline{quadratic} in the theta angle. So, let us turn to a second application with the ($p\pi^-$ loop-induced) neutron electric dipole moment \underline{linear} in $\theta_{\textrm{{\tiny $M$}}}$.

\subsection{Nucleon electric dipole moment}

Working in the isospin limit $m_u = m_d \equiv m_q$, the $T$-conserving effective interaction
\[
\Delta L_m (N) \ni \frac{b \ m_q}{f^2} \overline{N}\pi N\eta_0,
\eqno{(2.96)}
\]
derived this time from Eq. (2.59), leads to a \underline{scalar} (i.e., $T$-violating) coupling if, again, $\eta_0$ is replaced by its v.e.v. given in Eq. (2.85). Consequently, the full pion-nucleon interaction is now defined by
\[
L_{\pi\textrm{{\tiny $NN$}}} = - \frac{1}{\sqrt{2}} \overline{N}(g_{\pi\textrm{{\tiny $NN$}}} \ i \ \gamma_5 + g^\theta_{\ \pi\textrm{{\tiny $NN$}}}) \pi N
\eqno{(2.97)}
\]
with
\[
g_{\pi\textrm{{\tiny $NN$}}} \approx \sqrt{2} \frac{M_N}{f} \qquad \qquad \qquad \qquad \qquad
\eqno{(2.98)}
\]
and
\[
g^\theta_{\ \pi\textrm{{\tiny $NN$}}} \approx \frac{m_q}{\Delta m_q} \frac{\Delta M_{\textrm{{\tiny $N$}}}}{\sqrt{2}f} [\frac{m_0^{2}}{m_0^{2} + m_\pi^{2}}] \theta_{\textrm{{\tiny $M$}}},
\eqno{(2.99)}
\]
the $T$-conserving and $T$-violating effective couplings, respectively.  For particles with spin $\vec s$ moving in an electromagnetic field $(\vec{E}, \vec{B})$, the classical dipole interactions are described by
\[
H = -(d\vec{E} + \mu \vec{B}) \cdot \vec{s}.
\eqno{(2.100)}
\]
At this level the spin can be viewed as an intrinsic angular momentum such that its transformation laws under $T$ and $P$ are the same as for the magnetic field, but opposite to the ones for the electric field. Accordingly, only a magnetic moment is allowed  in any $T$-invariant theory.  The Dirac relativistic equation alone tells us that the electron should have a magnetic moment given by
\[
\mu_e = - \frac{e\hbar}{2m_e c}.
\eqno{(2.101)}
\]
Were the proton and neutron elementary particles, the Dirac theory would then also predict
\[
\mu_p^{\textrm{{\tiny ($D$)}}} = +  \frac{e\hbar}{2m_pc} \ , \ \mu_n^{\textrm{{\tiny ($D$)}}}=0.
\eqno{(2.102)}
\]
In fact, measurements yield anomalous magnetic moments:
\[
\mu_p \approx + 2.79 \mu_{\textrm{{\tiny $N$}}} \ , \ \mu_n \approx -1.91 \mu_{\textrm{{\tiny $N$}}}
\eqno{(2.103)}
\]
with 
\[
\mu_{\textrm{{\tiny $N$}}}  \equiv \frac{e\hbar}{2M_{\textrm{{\tiny $N$}}}   c} \approx 10^{-14} \textrm{e.cm.}
\eqno{(2.104)}
\]
the nuclear magneton (remember, $\hbar c \approx 197 \UMeV{}$ Fermi). These large departures from the predicted Dirac values are consequences of the fractional charge of the quarks confined in the nucleons. At the effective level, the magnetic  dipole moment of the neutron can be associated with its charged pion cloud. In this heuristic picture, the neutron electric dipole moment obtained by substituting $g^\theta_{\pi \textrm{{\tiny $NN$}}}$ for the left or right $g_{\pi \textrm{{\tiny $NN$}}}$ vertex of the associated diagram

\begin{center}
\includegraphics[bb=200 578 395 721]{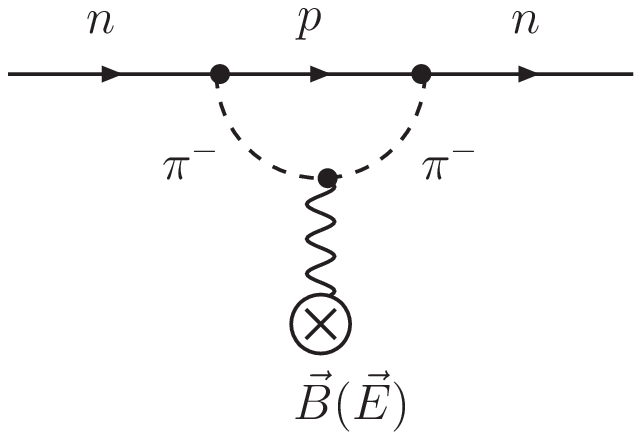}
\end{center}

\par\noindent
 is thus expected to scale like
\[
d_n \approx 2 [\frac{g^\theta_{\ \pi\textrm{{\tiny $NN$}}}}{g_{\pi\textrm{{\tiny $NN$}}}}] \mu_n.
\eqno{(2.105)}
\]
Neglecting again electromagnetic contributions to the proton-neutron mass difference, we obtain then
\[
d_n \approx [\frac{m_n-m_p}{m_n+m_p}] \cdot [\frac{m_d+m_u}{m_d-m_u}] \theta_{\textrm{{\tiny $M$}}} \ \mu_n.
\eqno{(2.106)}
\]
Compared with the present experimental limit,
\[
|d_n| < 2.9\times\!10^{-26} \textrm{e.cm.},
\eqno{(2.107)}
\]
the approximate expression (2.106) confirms the quite impressive bound first derived in \cite{17}:
\[
\theta_{\textrm{{\tiny $M$}}} < 10^{-9}.
\eqno{(2.108)}
\]
Such a strong constraint challenges theoreticians since decades. The  fine-tuning we face here for the time-reversal violation in a quantum theory of strong interactions (QCD) is rather similar to the fine-tuning for the vacuum energy density in a relativistic theory of gravitational interactions (GR):
\[
\theta \equiv  \arg \det m + \theta_{\textrm{{\tiny QCD}}} \approx 0 \Leftrightarrow (2\times\!10^{-3} \ \mbox{eV})^4 \approx \frac{c^2}{8\pi G} \Lambda_{\textrm{{\tiny $GR$}}} + \frac{f^2m^2_\pi}{2}  \equiv \rho_{\textrm{{\tiny vac.}}}.
\eqno{(2.109)}
\]
Two \textit{ad hoc} parameters, $\theta_{\textrm{{\tiny QCD}}}$ and $\Lambda_{\textrm{{\tiny $GR$}}}$, are indeed introduced by hand to reconcile our theoretical prejudices  with observations. Possible issues for the strong $\theta$-puzzle are in fact inspired by attempts to solve the cosmological $\Lambda$-problem. Let us briefly consider two of them.

\bigskip

\par\noindent
Firstly, by analogy with quintessence models which promote the cosmological constant $\Lambda$ at the level of a field, one may transform the $\theta$-parameter into a new dynamical variable 
\[
\theta(x) = \frac{2a_0 (x)}{F}.
\eqno{(2.110)}
\]
All the pseudoscalar fields have then a zero v.e.v. since
\[
<0|\eta_0|0> = <0|a_0|0 > =  0
\eqno{(2.111)}
\]
corresponds to the minimum of the new effective theory
\[
L(\eta_0 , a_0) = \frac{1}{2}  \partial_\mu \eta_0 \partial^\mu \eta_0 + \frac{1}{2}  \partial_\mu a_0\partial^\mu a_0 -  \frac{1}{2}  \frac{m_0^{\ 2}}{F^2} [F \eta_0 + f a_0]^2
\eqno{(2.112)}
\]
obtained after a field redefinition analogous to Eq. (2.81), i.e.,
\[
U \to \exp (\frac{ia_0}{F})U.
\eqno{(2.113)}
\]
As a consequence, the $T$ and $P$ discrete transformations are conserved in strong interactions but the spectrum of light pseudoscalars is modified. In the limit of massless quarks, the heavy iso-singlet pseudoscalar present in Eq. (2.112) is indeed given by 
\[
\eta' = \frac{[F\eta_0 + f a_0]}{(F^2+f^2)^{\frac{1}{2}}}
\eqno{(2.114)}
\]
while a new Goldstone boson, the axion, appears as the orthogonal combination:
\[
a =  \frac{[-f \eta_0 + F a_0]}{(F^2 + f^2)^{\frac{1}{2}}}. 
\eqno{(2.115)}
\]
The axion can be treated as the light brother of $\eta'$. Yet, despite an efficient $\Delta I = \frac{1}{2}$ contribution through its $\eta_0$ component which implies 
\[
\textrm{Br} \  (K^+ \to \pi^+ a) \approx \frac{f^2}{F^2} \textrm{Br} \  (K^0 \to \pi^+ \pi^-),
\eqno{(2.116)}
\]
it has never been seen so far and the present bound is
\[
\textrm{Br} \   (K^+ \to \pi^+ a) < 6\times\!10^{-11}.
\eqno{(2.117)}
\]
This direct limit from particle physics already puts a rather severe constraint on the scale $F$, namely
\[
F > 10^4  \UGeV{}.
\eqno{(2.118)}
\]
Consequently,  the  scale $F$  associated  with  the  spontaneous  symmetry  breaking   at  the  origin  of  the axion cannot be identified with the Fermi scale and the original Peccei-Quinn scenario \cite{18} is   excluded by this simple exercise.  Once the $u$ and $d$ quark masses are taken into account, the neutral squared mass matrix 
\[
m_{\textrm{\ {\tiny neutral}}}^2 = 
\begin{pmatrix}
 m_{\pi^\pm}^{2} & 0 & 0\\
 &&
\\
 0 & m_0^{2} + m_{\pi^\pm}^{2} & (\frac{f}{F}) m_0^{2}
 \\
 &&
 \\
 0 & (\frac{f}{F}) m_0^{2} & (\frac{f}{F})^2 m_0^{2}
\end{pmatrix}
\eqno{(2.119)}
\]
implies that the axion is in fact a pseudo-Goldstone boson with a mass given by
\[
m_a \approx  \frac{f}{F}  m_\pi < 1 \UkeV{}.
\eqno{(2.120)}
\]
Being a light cousin of the neutral pion, it only decays into two photons (in a $P$-wave) and its life-time scales like
\[
\tau (a \to \gamma\gamma) \approx (\frac{m_\pi}{m_a})^3 (\frac{F}{f})^2 \tau (\pi \to \gamma\gamma) \approx (\frac{F}{f})^5\times\!10^{-16} s.
\eqno{(2.121)}
\]
Axions couple to electromagnetic fields just as neutral pions do via the well-known  Primakoff effect. So, if axions exist, they should be produced at the solar core and immediately leave the Sun without further scattering, carrying an energy of the order of $T_{\textrm{\ {\tiny core}}}$ = 10$^7 K (\approx 1 \UkeV{})$. This has to be contrasted with  photons which scatter for about 10$^7$  years before reaching the surface of the Sun with an energy of the order of $T_{\textrm{\ {\tiny surface}}}$  = 6000 $K (\approx 1 \UeV{})$. From the known energy loss of the Sun, one infers the bound \cite{19}
\[
F > 10^7 \UGeV{}.
\eqno{(2.122)}
\]
This indirect astrophysical limit pushes the allowed lifetime of the axion far beyond the age of the Universe, promoting in this way the elusive particle at the level of a candidate for dark matter in cosmology if $F < 10^{12} \UGeV{}...$

\bigskip

\par\noindent
Secondly, by analogy with supersymmetry which ensures a vanishing vacuum energy, one may also impose an extra chiral symmetry which allows us to rotate away the $\theta$-parameter if
\[
 \det m=0.
 \eqno{(2.123)}
\]
However, the possibility of having a massless quark is hardly consistent with the isospin violation extracted from the mass spectrum of the full $0^{-+}$ nonet, i.e. 
\[
\frac{m_u}{m_d} \approx \frac{1}{2} \neq 0,
\eqno{(2.124)}
\]
and is even ruled out by large-$N_c$ arguments \cite{20}.

\par\noindent
As a matter of fact,   we now have to address the question of the origin of the quark (and lepton) masses beyond Newton's classical   definitions: 
\begin{enumerate}
\item[-] \underline{the measure of inertia} $(\frac{F}{a})$: a body tends indeed to resist any change in its existing state of rest or uniform motion, but the confinement of coloured particles tells us that quarks are never at rest and never free;
\item[-] \underline{the amount of matter} $(\rho V)$: it is obvious that an elephant weighs much more than a mouse because it is made of many more atoms than a mouse, but   elementary particles like the electron and the top quark which appear on an equal footing in quantum field theory obey the hierarchy
\[
\frac{m_{\textrm{{\tiny    electron}}}}{m_{\textrm{{\tiny  top}}}} 
\approx 
\frac{m_{\textrm{{\tiny mouse}}}}{m_{\textrm{{\tiny elephant}}}}.
\eqno{(2.125)}
\]
\end{enumerate}

\medskip

\par\noindent
In Section 1, we have seen that the bulk of our weight is due to the nucleon mass. Why should one then worry about the  electron mass? Well, the electron is the substance from which the chemical elements are built (see Mendeleev's Table). Its mass determines the size of atoms 
through the Bohr radius $(\div \  m^{\ -1}_e)$ or, to be more precise, the quantized energy levels with
\[
13.6\frac{\UeV{}}{c^2} = [\frac{\alpha^2}{2} + {\cal O} (\alpha^4)] m_e c^2
\eqno{(2.126)}
\]
for the hydrogen atom in Dirac's theory. So, no electron mass, no atoms but no atoms, no chemistry... Similarly, no (up and down) quark mass, no stable proton but no stable proton, no chemistry again!

\section{Spontaneous Symmetry Breaking and \mbox{\boldmath $[M_u,M_d] \neq 0$}}

\subsection{Boson masses and mixing}

Another way to bring the axial U(1) problem to an issue without introducing $T$-violation in the gauge theory for strong interactions is to assume the chiral symmetry breaking pattern
\[
\textrm{SU(2)}_{\textrm{{\tiny $L$}}} \times \textrm{SU(2)}_{\textrm{{\tiny $R$}}} \to \textrm{SU(2)}_{\textrm{{\tiny $V$}}}
\eqno{(3.1)}
\]
instead of (2.67).  If such was the case, only \underline{three} pseudoscalar Goldstone bosons would be produced out of an order parameter made of four degrees of freedom:
\[
\chi \equiv \frac{(\sigma +i\pi)}{\sqrt{2}}  \qquad  (\sigma = \sigma^0\tau_0 \ \ , \ \ \pi = \pi^a \tau_a) \ ,
\eqno{(3.2)}
\]
the missing $\eta_0$ would not trigger the axial U(1) problem and strong interactions would respect time-reversal symmetry. However, we know that a full decoupling of the pseudoscalar $\eta_0$ is not compatible with the  sizeable $\eta - \eta'$ mixing  given in Eq. (2.93).

\bigskip

\par\noindent
It turns out that the restricted chiral symmetry breaking (3.1) is quite relevant for the gauge theory of electroweak interactions. Indeed, the local invariance under SU(2)$_{\textrm{{\tiny $L$}}} \times$ U(1)$_{\textrm{{\tiny $Y$}}}$ of the Standard Model has to be spontaneously broken into U(1)$_{\textrm{{\tiny $Q$}}}$ with $Q$, the conserved electric charge:
\[
Q \equiv T_{3\textrm{{\tiny $L$}}} + \frac{Y}{2}.
\eqno{(3.3)}
\]
So,  a set of \underline{three} (eaten up) Goldstone bosons $(\pi = \pi^a \tau_a)$ is precisely what is needed to preserve one local U(1) unbroken and, therefore,  to guarantee that the zero photon mass is not a mere accident \cite{21} !

\bigskip

\par\noindent
Let us  again make use of the polar theorem (see Eq. (2.22)) to write
\[
\chi \equiv \xi(\pi)  \frac{\sigma}{\sqrt{2}}  \xi(\pi) =   \frac{\sigma^0}{\sqrt{2}}  U(\pi).
\eqno{(3.4)}
\]
 In the limit where the iso-singlet scalar field $\sigma^0$ is frozen at its v.e.v.,  
\[
< 0 | \sigma^0 | 0 > = v = (\sqrt{2} G_{\textrm{\ {\tiny Fermi}}})^{-\frac{1}{2}} \approx 246 \UGeV{},
\eqno{(3.5)}
\]
an iso-triplet of Goldstone fields $(\pi^\pm , \pi^3)$ is embodied in the unitary field 
\[
U = \exp (\frac{i\pi}{v})
\eqno{(3.6)}
\]
which globally transforms as
\[
U \to g_{\textrm{{\tiny $L$}}} Ug_{\textrm{{\tiny $R$}}}^{\dag}
\eqno{(3.7)}
\]
under SU(2)$_{\textrm{{\tiny $L$}}} \times$ SU(2)$_{\textrm{{\tiny $R$}}}$. The chiral  invariant kinetic term 
\[
L_{\textrm{\ {\tiny kinetic}}} (\pi) = \frac{v^2}{4} \ \textrm{Tr} \ (\partial_\mu U \partial^\mu U^\dag)
\eqno{(3.8)}
\]
analogous to Eq. (2.31) contains all the information about the scalar sector of the Standard Model, except of course for the elusive Higgs particle $(h = \sigma^0 - v)$.

\bigskip

\par\noindent
Gauging now the subgroup SU(2)$_{\textrm{{\tiny $L$}}} \times$ U(1)$_{\textrm{{\tiny $Y$}}}$ with the following normalizations
\[
T_{3\textrm{{\tiny $L$}}} = \frac{\tau_3}{2}
\eqno{(3.9)}
\]
\[   
\frac{Y}{2} = T_{3\textrm{{\tiny $R$}}} + \frac{B}{2}
\eqno{(3.10)}
\]
requires, as we know, the introduction of covariant derivatives. The baryon number $B$ is, by definition, vanishing for scalar fields. From the chiral transformations of $U$ in Eq. (3.7), we write therefore  a covariant derivative similar to Eq. (2.46):
\[
D_\mu U = \partial_\mu U - i  \frac{g}{2}  W^{\textrm{{\tiny $L$}}}_{\ \ \mu} U + i  \frac{g'}{2}  U W^{\textrm{{\tiny $R$}}}_{\ \ \mu}
\eqno{(3.11)}
\]
with
\[
W^{\textrm{{\tiny $L$}}}_{\ \ \mu} = 
\begin{pmatrix}
W_\mu^{\ \ 3} & \sqrt{2} W^{\ +}_\mu\\
 &&
\\
\sqrt{2} W^{\ -}_\mu & - W^{\ \ 3}_\mu
 \end{pmatrix}
\  \textrm{and} \ 
 W^{\textrm{{\tiny $R$}}}_{\ \ \mu} = 
\begin{pmatrix}
B_\mu  & 0 \\
 &&
\\
0 & - B_\mu
 \end{pmatrix}.
 \eqno{(3.12)}
\]
Note that the absence of charged gauge bosons in the $W^{\textrm{{\tiny $R$}}}_{\ \ \mu}$  matrix implies a (maximal) parity-violation through an explicit breaking of the SU(2)$_{\textrm{{\tiny $R$}}}$ global symmetry. Expanding the Goldstone field $U$ to zero order (or, equivalently, working in the unitary gauge $U=1\hspace{-0.7mm}\raisebox{0.5mm}{$\scriptstyle |$}$) in
\[
L_{\textrm{\ {\tiny non-linear}}} (\pi) = \frac{v^2}{4}  \ \textrm{Tr} \ (D_\mu U D^\mu U^\dag),
\eqno{(3.13)}
\]
we directly read the mass spectrum for the gauge bosons from
\[
\textrm{Tr} \ 
\begin{pmatrix}
(gW^{\ 3}_\mu - g' B_\mu) & g\sqrt{2}  W^{\ +}_\mu \\
&\\
g \sqrt{2}  W^{\ -}_\mu        & - (gW^{\ 3}_\mu - g' B_\mu) 
 \end{pmatrix}^2
 \frac{v^2}{16} \equiv \frac{1}{2} M^2_Z Z_\mu Z^\mu + M^2_W W^+_\mu W^{-\mu}.
 \eqno{(3.14)}
\]
The massive neutral gauge boson is the linear combination of $W^{\ 3}_\mu$ and $B_\mu$  present in this trace: 
\[
Z_\mu = \frac{(g W^{\ 3}_\mu - g' B_\mu)}{(g^2+g'^2)^{\frac{1}{2}}}
\eqno{(3.15)}
\]
with
\[
M_Z = (g^2 + g'^2)^{\frac{1}{2}} \frac{v}{2}.
\eqno{(3.16)}
\]
The orthogonal combination, absent from this trace, is naturally identified as the massless photon:
\[
V_\mu = \frac{(g' W^{\ 3}_\mu + g B_\mu)}{(g^2+g'^2)^{\frac{1}{2}}}.
\eqno{(3.17)}
\]
From the electromagnetic minimal substitution
\[
\partial_\mu U \to D_\mu U = \partial_\mu U -  i V_\mu [Q,U], \ \ Q = e (\frac{\tau_3}{2} + \frac{\tau_0}{6})
\eqno{(3.18)}
\]
already introduced in Eq. (2.52), we infer that
\[
e = \frac{gg'}{(g^2+g'^2)^{\frac{1}{2}}}.
\eqno{(3.19)}
\]
If we define the positron electric charge as
\[
e \equiv g \sin \theta_{\textrm{{\tiny $W$}}},
\eqno{(3.20)}
\]
then 
\[
\frac{g'}{g} = \tan \theta_{\textrm{{\tiny $W$}}}.
\eqno{(3.21)}
\]
Finally, the charged gauge bosons $W^{\ \pm}_\mu$ have a mass given by
\[
M _W = \frac{gv}{2} (\approx \frac{v}{3})
\eqno{(3.22)}
\]
such that a relation between physical quantities, i.e.,
\[
\frac{M^2_W}{M^2_Z}  = [\frac{g^2}{g^2+g'^2}] = \cos^2 \theta_{\textrm{{\tiny $W$}}}
\eqno{(3.23)}
\]
holds true, in remarkable agreement with current precision data.

\bigskip

\par\noindent
In the limit $g' \to 0$, the massive weak gauge bosons  $(W^{\pm}, W^{3})$ form another iso-triplet with respect to the global SU(2)$_{\textrm{{\tiny $V$}}}$ subgroup. Consequently, this unbroken hidden symmetry "protects" the tree-level $W - Z$ mass relation (3.23) against large radiative corrections. These one-loop quantum corrections grow logarithmically with the Higgs mass (which acts here as an ultraviolet cut-off), not quadratically. The "custodial"  symmetry being a successful feature of the Standard Model, it provides a rather severe constraint on any possible extension of its scalar sector. For illustration, a Two-Higgs-Doublet-Model (2HDM) characterized by a pair ($H^{\pm}$) of physical charged scalars should display some degeneracy in its   scalar mass spectrum. Indeed, yet another iso-triplet can be formed with either a CP-odd neutral scalar ($A^0$), as it is the case for the Minimal-Supersymmetric-Standard-Model (MSSM) in a decoupling limit:
\[
M^2_{H^\pm} = M^2_{A^0} + M^2_W \to M^2_{A^0},
\eqno{(3.24)}
\]
or with a CP-even one ($H^0$):
\[
M^2_{H^\pm} = M^2_{H^0},
\eqno{(3.25)}
\]
as it is the case if a twisted custodial symmetry is imposed \cite{22}. Note that new interesting LHC phenomenology may take place within the second scenario since the absence of a  $Z Z A^0$ coupling allows us to consider  $A^0$ as light as 50\UGeV{}~!

\subsection{Fermion masses, mixings and phase}

In the Standard Model for electroweak interactions, fermion masses are generated through arbitrary Yukawa interactions. \underline{If} the global SU(2)$_{\textrm{{\tiny $L$}}} \times$ SU(2)$_{\textrm{{\tiny $R$}}}$ symmetry of the scalar sector is extended to the fermions, we have
\[
L_{\textrm{\ {\tiny Yukawa}}}  = - Y_{ij} \overline{\Psi}^{0\ \ i}_{\ \textrm{{\tiny $L$}}} \chi \Psi^{0\ \ j}_{\ \textrm{{\tiny $R$}}} + \ \textrm{h.c.} \qquad  (i,j=1,...,N_g).
\eqno{(3.26)}
\]
The Latin indices assigned here to the (left and right-handed) quark doublets
\[
\Psi^0_{\textrm{{\tiny $L,R$}}} = 
\begin{pmatrix}
u^0\\
d^0
\end{pmatrix}_{\textrm{{\tiny $L,R$}}}
\eqno{(3.27)}
\]
run in the (three-dimensional) generation space. Once the order parameter is frozen at its SU(2)$_{\textrm{{\tiny $V$}}}$-inva\-riant v.e.v., 
\[
<0|\chi|0> =  \frac{1}{\sqrt{2}}  
\begin{pmatrix}
v & 0\\
0 & v
\end{pmatrix},
\eqno{(3.28)}
\]
we necessarily obtain equal mass matrices for the up and down quarks,
\[
M_{\textrm{\ {\tiny up}}}  = M_{\textrm{\ {\tiny down}}},
\eqno{(3.29)} 
\]
in a way similar to what happens for the nucleons (see Eq. (2.17)). But here only the heaviest quark appears to satisfy an approximate   Goldberger-Treiman   relation, 
\[
m_{\textrm{\ {\tiny top}}} \approx y_{tt}  \frac{v}{\sqrt{2}}, 
\eqno{(3.30)}
\]
with $y_{tt} \approx 1$. So, today the question is no more why is $t$ quark so heavy but why are the other quarks and leptons so light (see the elephant and the mouse...)? In the Standard Model, different Yukawa couplings to the right-handed quark fields are introduced to break the SU(2)$_{\textrm{{\tiny $V$}}}$ custodial symmetry. In other words, the invariance under the global SU(2)$_{\textrm{{\tiny $R$}}}$ is explicitly broken, as it was already the case when gauging the kinetic term for the Goldstone bosons via Eq. (3.11).  The order parameter $\chi$ transforms as
\[
\chi \to g_{\textrm{{\tiny $L$}}} \chi g_{\textrm{{\tiny $R$}}}^{\ \dag}
\eqno{(3.31)}
\]
under the global SU(2)$_{\textrm{{\tiny $L$}}} \times$ SU(2)$_{\textrm{{\tiny $R$}}}$. The local SU(2)$_{\textrm{{\tiny $L$}}}$ acting on $\chi$ from the left, let us write it in a bi-doublet matrix form 
\[
\chi = (H | -i \tau_2 H^\ast) = 
\begin{pmatrix}
\phi^0 & -(\phi^-)^\ast \\
\phi^- & (\phi^0)^\ast
\end{pmatrix}
\eqno{(3.32)}
\]
with the help of two complex fields
\[
\phi^0 =  \frac{1}{\sqrt{2}}  (\sigma^0 + i \pi^3)
\eqno{(3.33)}
\]
\[
\phi^- =  \frac{1}{\sqrt{2}}  (i \pi^1 - \pi^2).
\eqno{(3.34)}
\]
Indeed, if $H$ transforms as a doublet under SU(2), 
\[
H \to \exp (i\varepsilon^a \tau_a) H,
\eqno{(3.35)}
\]
so does $(-i \tau_2 H^\ast)$ since the Pauli matrices (2.8) satisfy the identities  
\[
(-i\tau_2) (-\tau_a^{\ \ast}) (-i \tau_2)^{-1} = \tau_a.
\eqno{(3.36)}
\]
Both the first column $(H)$ and the second column $(-i \tau_2 H^\ast)$ of $\chi$  transform as complex doublets under the local SU(2)$_{\textrm{{\tiny $L$}}} \times$ U(1)$_{\textrm{{\tiny $Y$}}}$, with hypercharge $Y= -1$ and $+1$ respectively. Accordingly, the most general gauge invariant Yukawa interactions are given by
\[
L_{\textrm{\ {\tiny Yukawa}}}  = - Y_{\textrm{{\tiny up}}}^{ij} \  \overline{\Psi}^0_{\ \textrm{{\tiny $Li$}}} \ H u^0_{\ \textrm{{\tiny $Rj$}}} - Y_{\textrm{{\tiny down}}}^{ij} \  \overline{\Psi}^0_{\ \textrm{{\tiny $Li$}}} (-i \tau_2 H^\ast) d^0_{\ \textrm{{\tiny $Rj$}}} + \ \textrm{h.c.}
\eqno{(3.37)}
\] 
In this way, the up and down quark mass matrices are unrelated and independent diagonalizations are required. Assuming (for a while!) hermitian mass matrices, we write
\[
M_{\textrm{{\tiny up}}}  =  Y_{\textrm{{\tiny up}}}  \frac{v}{\sqrt{2}}  = V^{\ \dag}_u D_u V_u
\eqno{(3.38)}
\]
\[
M_{\textrm{{\tiny down}}} =  Y_{\textrm{{\tiny down}}}  \frac{v}{\sqrt{2}}  = V^{\ \dag}_d D_d V_d
\eqno{(3.39)}
\]
with 
\[
D_u  =  \ \textrm{diag} \  (m_u, m_c, m_t)
\eqno{(3.40)}
\]
\[
D_d =  \ \textrm{diag} \  (m_d, m_s, m_b)
\eqno{(3.41)}
\]
and
\[
V_u V^{\ \dag}_u = V_d V^{\ \dag}_{d} = 1 \hspace{-0.7mm}\raisebox{0.5mm}{$\scriptstyle |$} .
\eqno{(3.42)}
\]
In the Standard Model for electroweak interactions, mixing angles arise from a misalignment between the gauge interaction basis $\{q^0\}$ and the mass matrix basis $\{q\}$ for the quark fields:
\[
\{q^0\} = V^\dag \{q\}.
\eqno{(3.43)}
\]
The left-handed charged current already introduced in Eq. (2.45) is now defined by 
\[
\bar{u}^{0\ i}_{\ \textrm{{\tiny $L$}}} \gamma^\mu \delta_{ij} d^{0\ j}_{\ \textrm{{\tiny $L$}}}  =  \bar{u}_{\textrm{{\tiny $L$}}}^{\ i} \gamma^\mu (V_{\textrm{{\tiny CKM}}})_{ij} d_{\textrm{{\tiny $L$}}}^{\ j}  
\eqno{(3.44)}
\]
and displays indeed a non-trivial Cabibbo-Kobayashi-Maskawa (CKM) mixing matrix 
\[
V_{\textrm{{\tiny CKM}}} \equiv V_u V_d^{\ \dag} \neq 1 \hspace{-0.7mm}\raisebox{0.5mm}{$\scriptstyle |$}
\eqno{(3.45)}
\]
whenever
\[
[M_u , M_d] \equiv  V_u^{\ \dag}  [D_{\textrm{{\tiny up}}}, V_{\textrm{{\tiny CKM}}} D_{\textrm{{\tiny down}}} V_{\textrm{{\tiny CKM}}}^{\ \ \dag}] V_u  \neq 0.
\eqno{(3.46)}
\]
Note that the trace of any power of this commutator, 
\[
\textrm{Tr} \  [M_u , M_d]^n =  \textrm{Tr} \  [D_{\textrm{{\tiny up}}}, V_{\textrm{{\tiny CKM}}} D_{\textrm{{\tiny down}}} V_{\textrm{{\tiny CKM}}}^{\ \ \dag}]^ n,
\eqno{(3.47)}
\]
defines an invariant which only depends on \underline{physical} quantities, namely  the quark masses in $D_{u,d}$, the mixing angles and, possibly, phases in $V_{\textrm{{\tiny CKM}}}$.

\bigskip

\par\noindent
    In the Standard Model for electroweak interactions, phases indeed arise from the arbitrary, i.e. complex, Yukawa couplings \cite{23}:
    \[
    Y_{\textrm{{\tiny up}}, \textrm{{\tiny down}}} \neq Y_{\textrm{{\tiny up}}, \textrm{{\tiny down}}}^{\ast}.
    \eqno{(3.48)}
    \]
Under the anti-unitary time-reversal operator (see Eq. (2.80)), each entry of the CKM mixing matrix is complex conjugated
\[
V_{\textrm{{\tiny CKM}}} \stackrel{{\textrm{{\tiny $T$}}}}{\to} V_{\textrm{{\tiny CKM}}}^{\ast}
\eqno{(3.49)}
\]
such that the invariant traces (3.47) transform as
\[
\textrm{Tr}\  [M_u , M_d]^n \stackrel{{\textrm{{\tiny $T$}}}}{\to}  (-1)^n \textrm{Tr}\  [M_u , M_d]^n.
\eqno{(3.50)}
\]
Consequently, we have a $T$-violation in weak interactions once
\[
\textrm{Tr}\  [M_u , M_d]^{2n+1} \neq 0 \ , \ n \geq 1.
\eqno{(3.51)}
\]
A well-known theorem named in honor of A. Cayley and W. Hamilton asserts that any $N \times N$ matrix $C$ is solution of its associated characteristic polynomial:
\[
p(\lambda) = \det (C -\lambda 1 \hspace{-0.7mm}\raisebox{0.5mm}{$\scriptstyle |$})  \Rightarrow p(C) = 0.
\eqno{(3.52)}
\]
Let us apply this theorem for the hermitian matrix
\[
C = i [ M_u , M_d].
\eqno{(3.53)}
\]

\noindent
\begin{enumerate}
\item[-] In the case of two generations $(N_g = 2)$, it simply implies that 
\[
p(C) = (C-c_1) (C-c_2) = C^2 - (\textrm{Tr} \ C)C + \det  C 1 \hspace{-0.7mm}\raisebox{0.5mm}{$\scriptstyle |$} = 0.
\eqno{(3.54)}
\]
The matrix $C$ being traceless, we have 
\[
C^2  = - \det C 1 \hspace{-0.7mm}\raisebox{0.5mm}{$\scriptstyle |$} .
\eqno{(3.55)}
\]
After $n$ iterations, we obtain
\[
\textrm{Tr} \  [M_u , M_d]^{2n+1} = (\det C)^n  \  \textrm{Tr} \ [M_u,M_d] = 0
\eqno{(3.56)}
\]
and time-reversal is always valid.

\bigskip

\item[-]   In the case of three generations $(N_g = 3)$,
\[
p(C) = (C-c_1) (C-c_2) (C-c_3) = C^3 - (\textrm{Tr}\  C) C^2 + \frac{1}{2} [(\textrm{Tr} \  C)^2 - \textrm{Tr} (C^2)] C - \det C 1 \hspace{-0.7mm}\raisebox{0.5mm}{$\scriptstyle |$} = 0
\eqno{(3.57)}
\]
and we have now 
\[
C^3 - \frac{1}{2} [\textrm{Tr} (C^2)] C - \det C 1 \hspace{-0.7mm}\raisebox{0.5mm}{$\scriptstyle |$} = 0.
\eqno{(3.58)}
\]
Taking then the trace, we conclude that
\[
\textrm{Tr} \  [M_u , M_d]^3  = 3 \det [M_u , M_d ] \neq 0
\eqno{(3.59)}
\]
and time-reversal is in principle violated.
\end{enumerate}

\par\noindent
Let us first consider a toy theory to illustrate a possible connection between mass-generation and $T$-violation. For that purpose, we simplify the flavour mixing pattern by assuming purely democratic transitions between the generations. In the two-generation case, it amounts to rotate the $d-s$ frame by a $\frac{\pi}{4}$ angle relative to the $u-c$ one and the Cabibbo mixing matrix is thus real: 
\[
V_{\textrm{{\tiny CKM}}} = (\sqrt{2})^{-1} 
\begin{pmatrix}
1 & 1 \\
-1 & 1
\end{pmatrix}.
\eqno{(3.60)}
\]
The generalization to the three-generation case is not obvious. Indeed, any $d-s-b$ frame rotation relative to the $u-c-t$ one violates democracy. We are therefore forced to work in a complex space with the introduction of a phase
\[
\omega = \exp (\frac{2i\pi}{3})
\eqno{(3.61)}
\]
 to guarantee full democracy in the moduli of the unitary CKM matrix \cite{24}:
\[
V_{\textrm{{\tiny CKM}}} = (\sqrt{3})^{-1} 
\begin{pmatrix}
1 & 1 & 1\\
\omega & 1 & \omega^2\\
\omega^2 & 1 &  \omega
\end{pmatrix}.
\eqno{(3.62)}
\]
This geometrical approach nicely confirms the previous mathematical theorem and exhibits the sharp difference between two and three generations of quarks as far as $T$-violation is concerned.

\bigskip

\par\noindent
Inspired by a rather successful mass relation in the charged lepton sector \cite{25},
\[
(m_e + m_\mu + m_\tau) = \frac{2}{3} \{\sqrt{m_e} + \sqrt{m_\mu} + \sqrt{m_\tau}\}^2,
\eqno{(3.63)}
\]
let us implement the CKM mixing matrix (3.62) with a down quark (hermitian) mass matrix
\[
M_{\textrm{{\tiny down}}} =  
\begin{pmatrix}
a          & b & b^\ast\\
b^\ast & a & b\\
b          & b^\ast &  a
\end{pmatrix}
\eqno{(3.64)}
\]
invariant under cyclic permutation ($d_3$ discrete group) in the basis where the up quark mass matrix is diagonal:
\[
M_{\textrm{{\tiny up}}} =  \textrm{diag} \  (m_u, m_c, m_t).
\eqno{(3.65)}
\]
Here, the misalignment given in Eq. (3.43) is due to impossibility of simultaneous diagonalization since the commutator of the two mass matrices reads
\[
[M_u , M_d] = 
 \begin{pmatrix}
0                             & b(m_c-m_u) & b^\ast(m_t-m_u)\\
-b^\ast(m_c-m_u) & 0 & b(m_t-m_c)\\
-b (m_t-m_u)         & -b^\ast(m_t-m_c) &  0
\end{pmatrix} \neq 0.
\eqno{(3.66)}
\]
Moreover, a $T$-violation occurs in this simple ansatz with democratic mixings since  
\[
\det  [M_u , M_d]  = (m_t - m_c) (m_t - m_u) (m_c - m_u) (b^{\ast 3} - b^3) \neq 0.
\eqno{(3.67)}
\]
The eigenvalues of the down mass matrix (3.64) are extracted from the relation
\[
M_{\textrm{{\tiny down}}} V_{\textrm{{\tiny CKM}}} = V_{\textrm{{\tiny CKM}}} D_{\textrm{{\tiny down}}}
\eqno{(3.68)}
\]
and are given by
\[
\begin{array}{lllll}
 m_d = a + b\omega + b^\ast \omega^2
\\
 m_s = a + b + b^\ast
\\
 m_b = a + b\omega^2 + b^\ast \omega.
\end{array}
\eqno{(3.69)}
\]
Consequently,
\[
(b^{\ast 3} - b^3) = \frac{2i}{9} (m_b - m_s)(m_b - m_d)(m_s - m_d)  \Im (\omega^2)
\eqno{(3.70)}
\]
and the determinant of the $[M_u , M_d]$  commutator only depends on \underline{physical} quantities (the quark masses in $D_{u,d}$ and the phase in $V_{\textrm{{\tiny CKM}}}$), as anticipated in Eq. (3.47).

\bigskip

\par\noindent
A way to restore $T$-invariance in this toy theory is to impose some mass degeneracy. For example, in the limit $b = b^\ast$, the down mass matrix (3.64) is invariant under permutations ($S_3$ discrete group) and admits two degenerate eigenvalues $(m_d = m_b)$.  In that limit the matrix is real and, consequently, $T$-conserving. Equivalently, a pseudo-rotation of 45$^\circ$ in the $d-b$ plane allows us to rotate away the $\omega$ phase and to write the CKM unitary matrix (3.62) as a "tri-bimaximal" orthogonal one: 
\[
V_{\textrm{{\tiny CKM}}}   
\begin{pmatrix} \displaystyle
\frac{1}{\sqrt{2}} & \displaystyle 0 & \displaystyle \frac{-i}{\sqrt{2}}\\
&&
\\
0 & 1 & 0
\\
&&
\\ \displaystyle
\frac{1}{\sqrt{2}} & 0 &  \displaystyle\frac{i}{\sqrt{2}}
\end{pmatrix}
=
\begin{pmatrix} \displaystyle
\frac{2}{\sqrt{6}} & \displaystyle \frac{1}{\sqrt{3}} & 0\\
&&
\\ \displaystyle
\frac{-1}{\sqrt{6}} & \displaystyle \frac{1}{\sqrt{3}}  & \displaystyle\frac{1}{\sqrt{2}} \\
&&
\\ \displaystyle
\frac{-1}{\sqrt{6}} & \displaystyle \frac{1}{\sqrt{3}}  & \displaystyle \frac{-1}{\sqrt{2}} 
\end{pmatrix}
\eqno{(3.71)}
\]
which appears to be of some relevance for neutrino physics.

\bigskip

\par\noindent
Our simple ansatz defined by  (3.64) and (3.65) for the quark mass matrices has revealed a deep connection between mass-splitting and $T$-violation. It also provides a rather easy way to understand the concepts of "unitarity triangles" and "$J$-invariant", respectively.

\bigskip

\par\noindent
In general, for three generations, six independent unitarity triangles ($UT$'s) in the complex plane are expected from the unitarity constraints 
\[
\sum_j (V_{\textrm{{\tiny CKM}}}^{\ \ \dag})_{ij}  (V_{\textrm{{\tiny CKM}}})_{jk}   = 0 \  \ \ \mbox{if} \  \ i\neq k .
\eqno{(3.72)}
\]
 In our toy theory (3.62), they reduce to a single equilateral $UT$ defined by the relation
\[
1 + \omega + \omega^2 = 0,
\eqno{(3.73)}
\]

\begin{center}
\psfrag{un}{\huge{UT}}
\psfrag{om2}{$\dfrac{\omega^2}{3}$}
  \psfrag{om3}{$\dfrac{\omega}{3}$}
  \psfrag{onethird}{$\dfrac{1}{3}$}

  \includegraphics[]{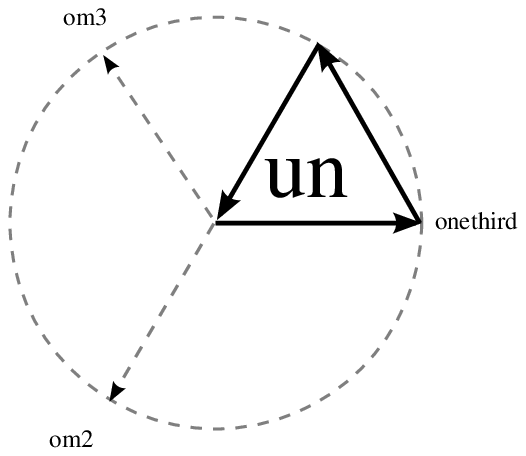}
\end{center}

\par\noindent
In general, for three generations, the $T$-violating invariant is defined by
\[
\det [M_u , M_d] = 2i (m_t-m_c) (m_t-m_u) (m_c-m_u) (m_b-m_s) (m_b-m_d) (m_s-m_d) J
\eqno{(3.74)}
\]
with
\[
J\equiv \pm \Im [(V)_{ij} (V^\dag)_{jk} (V)_{kl} (V^\dag)_{li}].
\eqno{(3.75)}
\]
As a mnemotechnic, you may consider the flavour structure of a quark loop with 2 virtual $W$'s:

\begin{center}
\includegraphics[keepaspectratio,height=6cm]{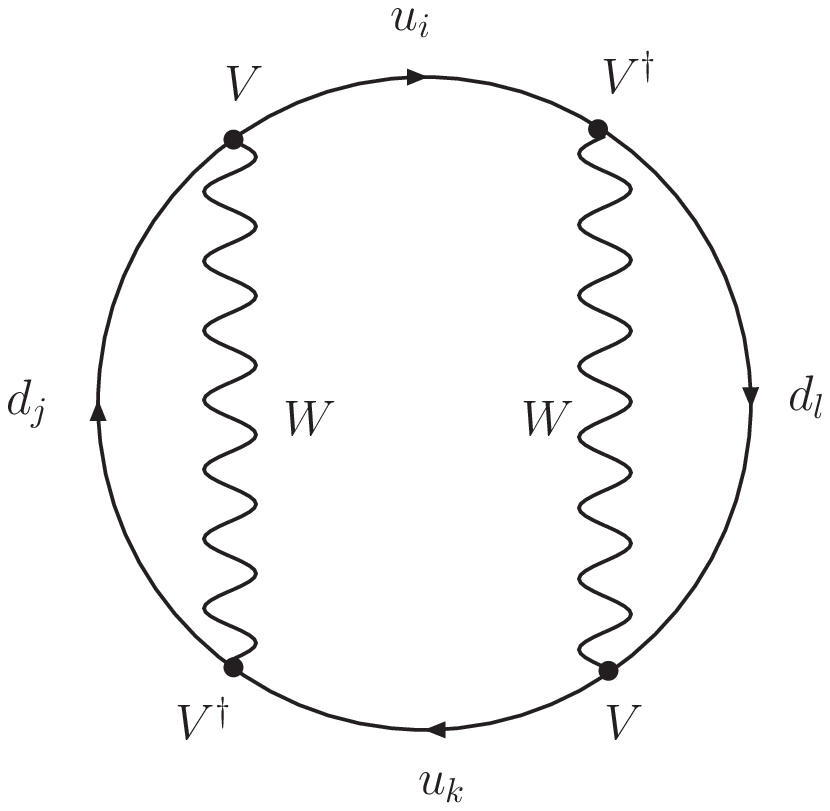}
\end{center}

\par\noindent
The imaginary part of all possible "quartet" $(V)_{ij} (V^\dag)_{jk} (V)_{kl} (V^\dag)_{li}$ (no sum over flavour indices!) being equal up to a sign, the absolute value of $J$ is unique and, in fact, proportional to the area $A$ of any $UT$:
\[
|J|=2A_\Delta.
\eqno{(3.76)}
\]
In our toy theory (3.62), we obtain indeed
\[
J \equiv \Im (V_{12} V_{22}^{\ \ \ast} V_{23} V_{13}^{\ \ \ast}) =  \frac{1}{9}  \Im (\omega^2) = \frac{-1}{6\sqrt{3}}.
\eqno{(3.77)}
\]
A more realistic CKM matrix has of course to be considered to reproduce the full $ (K^0 , B^0)$ phenomeno\-logy. Expanding in the Cabibbo angle
\[
\theta_c = \lambda \approx 0.23,
\eqno{(3.78)}
\]
one gets the following (very rough) pattern for the flavour mixings
\[
 \begin{pmatrix}
   \cos \theta_c   & \sin \theta_c  \\
    -\sin \theta_c  & \cos \theta_c
\end{pmatrix}
\stackrel{C}{\approx}
\begin{pmatrix}
1 & \lambda
\\
-\lambda & 1
 \end{pmatrix}
 \stackrel{KM}{\to}
 \begin{pmatrix}
   1   &  \lambda & -\lambda^3 \omega  \\
    -\lambda  & 1 & \lambda^2\\
    -\lambda^3 \omega & -\lambda^2 & 1
\end{pmatrix}.
\eqno{(3.79)}
\]
At this level of approximation, we consider
\begin{enumerate}
\item[-] one $UT$ directly accessible at $\mathcal{O}(\lambda^3)$ in $B_d$-physics:
 \[
 (V^\dag V)_{db} = V_{ud}^{\ \ \ast} V_{ub} + V_{cd}^{\ \ \ast} V_{cb} + V_{td}^{\ \ \ast} V_{tb} \approx - \lambda^3(\omega^2+1+\omega) = 0
 \eqno{(3.80)}
 \]
 \item[-] the invariant
 \[
 J = \Im (V_{12} V_{22}^{\ \ \ast} V_{23} V_{13}^{\ \ \ast}) \approx \lambda^6 \Im (\omega)  \approx 10^{-4}.
 \eqno{(3.81)}
 \]
\end{enumerate}
The hierarchy observed in the CKM mixing matrix may suggest that the phenomenon of flavour mixing is intimately related to the quark mass spectrum. Specific textures have indeed been proposed in that context. For illustration, the two-by-two quark mass matrix
\[
M_{\textrm{{\tiny down}}} =  
\begin{pmatrix}
0          & b \\
b          &   a
\end{pmatrix}
\eqno{(3.82)}
\]
implies an intriguing relation between the Cabibbo mixing angle and a mass ratio \cite{26},
\[
\lambda \approx (\frac{m_d}{m_s})^{\frac{1}{2}},
\eqno{(3.83)}
\]
which triggered so many attempts to derive the CKM matrix (3.79) from "horizontal" symmetries acting in the generation space. The large value conventionally  extracted for the CKM phase may rather suggest a "geometrical" $T$-violation with \cite{27}  
\[
\delta_{\textrm{{\tiny CKM}}} = \frac{2\pi}{N_g}
\eqno{(3.84)}
\]
only depending on the number $N_g$ of generations, not on a mass ratio. However, one should keep in mind that there are in fact nine distinctive parametrizations $(P_i)$ of the CKM matrix,

\[
\begin{pmatrix} 
\ast & \ast & \circ \\
\ast & \ast & \circ \\
\circ & \circ & \circ 
\end{pmatrix}
\quad
\begin{pmatrix} 
\ast & \circ & \ast \\
\circ & \circ & \circ \\
\ast & \circ & \ast 
\end{pmatrix}
\quad
\begin{pmatrix} 
\circ & \circ & \circ \\
 \circ &\ast & \ast \\
\circ & \ast  & \ast 
\end{pmatrix}
\]
\[
\  P_1 \hspace{21mm} P_2 \hspace{20mm} P_3
\]
\[
\begin{pmatrix} 
\circ & \ast & \ast  \\
\circ & \ast & \ast  \\
\circ & \circ & \circ 
\end{pmatrix}
\quad
\begin{pmatrix} 
\circ & \circ & \circ \\
\ast & \ast & \circ \\
\ast & \ast & \circ
\end{pmatrix}
\quad
\begin{pmatrix} 
\ast & \circ &    \ast \\
\ast & \circ &    \ast \\
\circ & \circ & \circ 
\end{pmatrix}
\]
\[
\  P_4 \hspace{21mm} P_5 \hspace{20mm} P_6
\]
\[ \quad
\begin{pmatrix} 
\circ & \circ & \circ \\
\ast & \circ &    \ast  \\
\ast & \circ &    \ast  
\end{pmatrix}
\quad
\begin{pmatrix} 
\ast & \ast & \circ \\
\circ & \circ & \circ \\
\ast & \ast & \circ
\end{pmatrix}
\quad
\begin{pmatrix} 
\circ & \ast & \ast \\
\circ & \circ & \circ  \\
\circ & \ast & \ast
\end{pmatrix}  \ \ ,
\]
\[
\  P_7 \hspace{21mm} P_8 \hspace{20mm} P_9
\]
each of them being obtained by imposing one row and one column to be real (see the circles in $P_i$).  The invariant quantity $J$ \cite{28} is indeed expressed in terms of four mixing matrix elements which always form a "plaquette" (see the asteriks in $P_i$). So, there are in principle nine independent phase conventions. For illustrations, the original KM parametrization corresponds to excluding the first row and the first column $(P_3)$ for the phase, while the standard convention is equivalent to crossing out the first row and the third column $(P_5)$. However, any fundamental theory hidden behind the observed hierarchical quark mass spectrum should privilege one of these nine parametrizations. In this respect, we note that one and only one of them allows a small phase. By crossing out the second row and the second column $(P_2)$, we obtain indeed $V_{ub} \sim \lambda (e^{-i\delta}-1)$ with  a $T$-violating angle $\delta$ of the order of 1$^\circ$:
\[
\delta (P_2)=(1.1 \pm 0.1)^\circ.
\eqno{(3.85)}
\]
Within this parametrization $P_2$, the three angles are roughly equal to $\lambda$ and the smallness of the $J$-invariant is accounted for by the smallness of the phase $(J \approx \lambda^4\delta)$.
A natural relation between the CKM phase and a mass ratio  is therefore also possible \cite{29}. For example, a "hierarchical" T-violation with 
\[
\delta_{\textrm{{\tiny CKM}}} = \frac{m_s}{m_b}
\eqno{(3.86)}
\]
consistently disappears in the decoupling limit $(m_b \to \infty)$ for the third generation.

\subsection{Matter/antimatter asymmetry}

In the Standard Model for electroweak interactions, Eq. (3.37) implies that all the presently observed $T$ (or $CP$-) violating phenomena originate from the complex ($CPT$-invariant) Yukawa couplings $Y$ of the Higgs field $h$ to the quarks since
\[
L_Y^{\ \ \textrm{{\tiny neutral}}} = -  \frac{1}{\sqrt{2}}  (\bar{q}^0_{\ \textrm{{\tiny $L$}}}  Y q^0_{\ \textrm{{\tiny $R$}}} + \bar{q}^0_{\ \textrm{{\tiny $R$}}} Y^\dag q^0_{\ \textrm{{\tiny $L$}}} ) (h + v)
\eqno{(3.87)}
\]
with
\[
M = Y  \frac{v}{\sqrt{2}}  \stackrel{T}{\to} M^\ast.
\eqno{(3.88)}
\]
In that sense, $CP$-violation is our second compelling argument in favour of a single Higgs field, the first one being the custodial symmetry at the source of a natural zero photon mass. A Multi-Higgs-Doublet-Model generically produces a massive photon as well as a large neutron electric dipole moment.

\bigskip

\par\noindent
However, there is no natural way to guarantee hermitian quark mass matrices in this Standard Model. It is a fact that the mass matrices transform as                                       
\[
M = Y  \frac{v}{\sqrt{2}}  \stackrel{P}{\to} M^\dag
\eqno{(3.89)}
\]
under parity which interchanges left-handed and right-handed fields in Eq. (3.87). But we cannot impose this discrete symmetry on the whole theory since the weak gauge interactions are known to violate parity. So we have to consider another invariant involving now the commutator of the $MM^\dag$ hermitian matrices:
\[
\begin{array}{cccc}
 \det [M_u M_u^{\ \dag} , M_d M_d^{\ \dag}] \equiv \det [D^2_{\ \textrm{{\tiny up}}} , V D^2_{\ \textrm{{\tiny down}}} V^\dag] &
 \\
 &
\\
 = 2i (m_t^{\ 2} - m_c^{\ 2}) (m_t^{\ 2} - m_u^{\ 2}) (m_c^{\ 2} - m_u^{\ 2}) (m_b^{\ 2} - m_s^{\ 2}) (m_b^{\ 2} - m_d^{\ 2}) (m_s^{\ 2} - m_d^{\ 2}) J. &
 \end{array}
 \eqno{(3.90)}
\]
It also contains all the information about $T$ violation (\textit{squared} mass splittings and phase) since
\[
\det [M_u M_u^{\ \dag} , M_d M_d^{\ \dag}] \stackrel{T}{\to}  - \det [M_u M_u^{\ \dag} , M_d M_d^{\ \dag}]
\eqno{(3.91)}
\]
but has clearly no defined parity:
\[
\det [M_u M_u^{\ \dag} , M_d M_d^{\ \dag}] \stackrel{P}{\to}  + \det [M_u^{\ \dag} M_u  , M_d^{\ \dag} M_d]
\eqno{(3.92)}
\]
as expected. So, let us invoke once more the polar decomposition, to write 
\[
M_{u(d)} = H_{u(d)} U^{u(d)}_{\textrm{{\tiny $R$}}}.
\eqno{(3.93)}
\]
A diagonalization in two steps is then necessary to bring both mass matrices into their diagonal, real form defining the physical quark states. First, one exploits the fact that the right-handed quark fields are sterile with respect to the charged weak currents to eliminate $U_{\textrm{{\tiny $R$}}}$ through a \underline{chiral} transformation. In this first step, the QCD axial anomaly in Eq. (2.76) just holds back the flavour singlet phase with angle
\[
\theta_{\textrm{{\tiny $M$}}} = \arg \det (U^u_{\textrm{{\tiny $R$}}}   U^d_{\textrm{{\tiny $R$}}}) \neq 0.
\eqno{(3.94)}
\]
The second step consists then in a \underline{vectorial} transformation acting equally on the left- and right-handed up (down) quark fields to diagonalize the remaining hermitian matrices $H_{u(d)}$. The observed mass hierarchy for the up and down quarks gives then
\[
2i  m_t^{\  2} m_c m_b^{\  2} m_s J \approx \det [H_u , H_d] \neq 0.
\eqno{(3.95)}
\]
In other words, the maximal parity-violation in the quark charged currents (3.44) implies that the $\theta_{\textrm{{\tiny $M$}}}$ and  $\delta_{\textrm{{\tiny CKM}}}$ angles are totally unrelated at the tree-level. On the one hand, strong $CP$ violation in  flavour diagonal transitions occurs through a $T$-violating quantity which is  $C$-even but $P$-odd since
\[
\arg \det (U^u_{\textrm{{\tiny $R$}}} U^d_{\textrm{{\tiny $R$}}}) \stackrel{P}{\to}  - \arg \det (U^u_{\textrm{{\tiny $R$}}}  U^d_{\textrm{{\tiny $R$}}}).
\eqno{(3.96)}
\]
And this is precisely what is required to generate an electric dipole moment for the neutron, as already displayed in Eq. (2.106):
\[
d_n \approx \theta\times\!10^{-16} \ \textrm{e.cm}.
\eqno{(3.97)}
\]
On the other hand, weak $CP$ violation in $(V-A)$ flavour changing transitions occurs through a $T$-violating quantity which is $P$-even but $C$-odd since  
\[
\det [H_u , H_d]  \stackrel{C}{\to}   -\det [H_u , H_d].
\eqno{(3.98)}
\]
And this is precisely one of the necessary ingredients to dynamically generate the matter/antimatter asymmetry observed in the present Universe \cite{29}:
\[
\frac{(n_{\textrm{{\tiny $B$}}} - n_{\overline{\textrm{{\tiny $B$}}}})}{n_\gamma} \Bigl|_0 = (6.1 \pm 0.2)\times\!10^{-10}.
\eqno{(3.99)}
\]
From the magnitude \underline{and} the quantum number assignment of its two independent sources of $T$-violation:
\[
\begin{array}{cccc}
 | \theta| < 10^{-9}  \ \  ; \ \ J^{\textrm{{\tiny $PC$}}} = 0^{-+}  &
 \\
 &
\\\displaystyle
 2 m_t^{\ 2} m_c m_b^{\ 2} m_s \frac{| J|}{(\frac{v}{\sqrt{2}})^6} \approx 10^{-14}      \ \ ; \ \ J^{\textrm{{\tiny $PC$}}} = 0^{+-}     &
 \end{array}
 \eqno{(3.100)}
\]
we conclude that the Standard Model for strong and electroweak interactions does not seem to be able to produce enough baryon asymmetry. However, both sources are deeply connected to the quark mass spectrum : the former vanishes if one quark is massless (say, $m_u = 0$), while the latter can be rotated away if two quarks with same electric charge are degenerated (say, $m_d = m_s$). So one may conjecture that they have in fact the same magnitude. If such turns out to be the case, $|\theta| \approx 10^{-14}$ and one expects a neutron electric dipole moment around $10^{-30}$ e.cm.

\section*{Conclusions}

Gauge invariance and time-reversal provide us with some (modest) steps towards a possible unification of the fundamental interactions. These symmetries explain for example why the weakest among the four known basic forces of nature, i.e. gravity, eventually dominates in the celestial environment (from the spherical shape of planets to the expansion of the Universe...). 
\begin{itemize}
\item Electromagnetic and gravitational interactions indeed obey gauge invariance which requires  massless messangers; so they both lead to long range forces. However, time-reversal applied on the corresponding connections ($-iA$ and $\Gamma$, respectively) disentangles them: screening   only occurs for spin 1 mediated interactions between opposite sign charges, not for spin 0 or 2 ones which couple positive masses or energies. 
\item Strong and weak interactions get round gauge invariance through the subtle mechanisms of confinement and spontaneous symmetry breaking, respectively. In this  way, $T$-violation is peculiar to short range nuclear forces. 
\end{itemize}

\par\noindent
This striking correlation between gauge invariance and time reversal symmetry challenges us. Questions at issue are the unexpectedly tiny value of the cosmological constant $\Lambda$ in the Einstein-Hilbert action and of the angle $\theta$ in the QCD action. A direct observation of non-baryonic dark matter and of the neutron electric dipole moment could bring these fundamental questions to a successful issue.

\bigskip

\par\noindent
Needless to emphasize that a theoretical understanding of the full fermion mass spectrum or (and) the discovery of the Higgs boson would be a major breakthrough in any case.

\begin{itemize}
\item If the Higgs boson turns out to be elementary, it will open the door to other hypothetical scalar fields (Quintessence, Inflaton or Axion...) invoked to solve further theoretical puzzles (dark ener\-gy, homogeneity, electric dipole moments...) in cosmology and particle physics. Moreover, its Yukawa interactions which are genuine sources for $T$-violation would be promoted at the rank of the 5$^{\textrm{th}}$ fundamental interaction and the issue of universal coupling reopened. Our knowledge about the gravitational interactions may help us in that venture. At this point we simply note that in the historical Thomson experiment which led to the discovery of the first elementary particle, only charged particles could feel the electric field, not neutral ones. Similarly, in the early Nordstr\"om's theory, only massive particles could feel the gravitational field, not massless ones. So, the Higgs boson in its present formulation looks more like a scalar graviton. Could the analogy with a more successful background-independent theory of gravity guide us towards a geometrical interpretation of the Yukawa interactions? 

\item If the Higgs boson is proving not elementary, no doubt that the strong interactions will continue to inspire us in the quest for our precise weight. After all, less than 5\% of the matter-energy content of our Universe is presently understood!

\end{itemize}

\section*{Acknowledgments}

I would like to thank C\'eline Degrande, Fabio Maltoni and Jacques Weyers for their comments on the manuscript as well as to Cathy Brichard and Vincent Boucher for their help in preparing these lecture-notes. This work was supported by the Belgian Federal Office for Scientific, Technical and Cultural Affairs through the Interuniversity Attraction Pole No. P6/11.

\end{document}

%% file: wasyfont.tex
\font\tenwasy = wasy10

\font\sevenwasy = wasy7
\font\fivewasy = wasy5
\newfam\wasyfam
\newcount\wasyfamcount
\wasyfamcount=\wasyfam \multiply\wasyfamcount by 256
\def\wasy{\fam\wasyfam\tenwasy}
\textfont\wasyfam=\tenwasy
\scriptfont\wasyfam=\sevenwasy
\scriptscriptfont\wasyfam=\fivewasy
\def\overstrike#1#2{{\setbox0\hbox{$#2$}\hbox to \wd0{\hss
    $#1$\hss}\kern-\wd0\box0}}

\def\AC{\hbox{\kern0.5pt\wasy\char"3A\kern0.5pt}}
\def\HF{\lower0.9pt\hbox to 0pt{\kern0.5pt\wasy\char"3A\hss}%
        \raise0.9pt\hbox{\kern0.5pt\wasy\char"3A\kern0.5pt}}

\def\LEFTcircle{\hbox to 0pt{\wasy\char"47\hss}\hbox{\wasy\char"23}}
\def\RIGHTcircle{\hbox to 0pt{\wasy\char"48\hss}\hbox{\wasy\char"23}}

\def\leftmoon{\hbox{\wasy\char"24}}

\def\venus{\raise0.2ex\hbox{\wasy\char"19}}
\def\earth{\lower0.3ex\hbox{\wasy\char"26}}
\def\mars{\lower0.2ex\hbox{\wasy\char"1A}}

\def\APLbox{{\hbox{\wasy\char"7E}}}

\def\APLinv{{\hbox to 0pt{\tensy\char"04\hss}\APLbox}}

\def\APLminus{\raise0.7ex\hbox{$-$}}

\def\Dh{\leavevmode{\rm\setbox0\hbox{D}%
    \hbox to\wd0{\kern 0.04em\char32\hss D}}}

\def\wasymathchardef#1#2{%
    \count255=\wasyfamcount
    \advance\count255 by"#1
    \mathchardef#2\count255}
\wasymathchardef{3001}{\lhd}
\wasymathchardef{3003}{\rhd}
\wasymathchardef{3010}{\LHD}
\wasymathchardef{3011}{\RHD}
\wasymathchardef{3002}{\unlhd}
\wasymathchardef{3004}{\unrhd}
\wasymathchardef{303C}{\sqsubset}
\wasymathchardef{303D}{\sqsupset}
\wasymathchardef{303E}{\apprle}
\wasymathchardef{303F}{\apprge}
\wasymathchardef{301D}{\varpropto}
\wasymathchardef{0018}{\invneg}
\wasymathchardef{303B}{\leadsto}
\wasymathchardef{2023}{\ocircle}
\wasymathchardef{3016}{\logof}
\wasymathchardef{1072}{\varint}
\wasymathchardef{1073}{\iint}
\wasymathchardef{1074}{\iiint}
\wasymathchardef{1075}{\varoint}
\wasymathchardef{1076}{\oiint}
\def\newpropto{\let\propto\varpropto}
\def\newint{\let\int\varint \let\oint\varoint} 